\documentstyle[aps,preprint,tighten,floats]{revtex}

\newcommand{\lsim}{\mathrel{\mathop{\kern 0pt \rlap
  {\raise.2ex\hbox{$<$}}}
  \lower.9ex\hbox{\kern-.190em $\sim$}}}
\newcommand{\gsim}{\mathrel{\mathop{\kern 0pt \rlap
  {\raise.2ex\hbox{$>$}}}
  \lower.9ex\hbox{\kern-.190em $\sim$}}}

\begin{document}

\preprint{
\begin{tabular}{r}
DFTT 14/98\\
LAPTH 676/98
\end{tabular}}

\title {\bf Which fraction of the measured cosmic--ray antiprotons might be due 
          to neutralino annihilation in the galactic halo?}

\author{
\bf A. Bottino$^{\mbox{a}}$
\footnote{E--mail: bottino@to.infn.it, donato@to.infn.it,
fornengo@to.infn.it, salati@lapp.in2p3.fr},
F. Donato$^{\mbox{a}}$, N. Fornengo$^{\mbox{a}}$,
P. Salati$^{\mbox{b}}$
\vspace{6mm}
}

\address{
\begin{tabular}{c}
$^{\mbox{a}}$
Dipartimento di Fisica Teorica, Universit\`a di Torino and \\
INFN, Sezione di Torino, Via P. Giuria 1, 10125 Torino, Italy
\\
$^{\mbox{b}}$Laboratoire de Physique Th\'eorique ENSLAPP, BP110, F--74941\\
Annecy--le--Vieux Cedex, France.
\end{tabular}
}
%\date{April 14, 1998}
\maketitle

\begin{abstract}
We analyze the data of low--energy cosmic--ray $\bar p$ spectrum, recently 
published by  the BESS Collaboration, in terms of newly calculated fluxes for
secondary antiprotons and for a possible contribution of an exotic signal due
to neutralino annihilation in the galactic halo. We single out the relevant 
supersymmetric configurations and discuss their explorability with experiments
of direct search for particle dark matter and at accelerators. We discuss how 
future measurements with the Alpha Magnetic Spectrometer (AMS) on 
the Shuttle flight may disentangle the possible 
neutralino--induced contribution from the secondary one. 
\end{abstract}

\section{Introduction}

A recent analysis \cite{bess95}
of the data collected by the balloon--borne BESS
spectrometer on cosmic--ray antiprotons during its flight in 1995 (hereafter 
referred to as BESS95 data)
has provided the most detailed information on the 
low--energy 
cosmic--ray $\bar{p}$'s spectrum currently available: 
43 antiprotons have been detected,
grouped in 5 narrow energy windows over the total kinetic--energy range 
$180 ~ {\rm MeV} \leq  T_{\bar{p}} \leq 1.4 ~{\rm GeV}$. With this 
experiment the total number of 
measured cosmic--ray antiprotons in balloon--borne detectors over a period of 
more
than 20 years \cite{all,hof,mitchell,moiseev,barbiellini} has more than 
doubled. Most remarkably, the BESS95 data
provide a very useful information over the low--energy part of the ${\bar p}$ 
flux,
where a possible distortion of the spectrum expected for secondary $\bar{p}$'s 
(i.e., antiprotons created by interactions of primary cosmic--ray nuclei
with the interstellar medium) may reveal the existence of cosmic--ray 
antiprotons of exotic origin (for instance, due to pair annihilation of relic
particles in the 
galactic halo \cite{th,noi,mitsui}, to evaporation of 
primordial black holes \cite{mitsui,kww} or to cosmic strings \cite{strings}). 
In fact, a possible discrimination between
primary (exotic) and secondary $\bar p$'s is based on the different features of
their low--energy spectra: in this energy regime 
($ T_{\bar p} \lsim $ 1 GeV)  interstellar (IS) secondary $\bar p$ spectrum 
is expected to drop off very markedly because of kinematical reasons 
\cite{gaisser}, while exotic antiprotons show a milder fall off. 
However, as will be discussed later on, this discrimination  power is 
somewhat hindered by solar modulation and by some other effects affecting 
particle diffusion in the Galaxy. 

In Fig. 1 we report the cosmic--ray $\bar p$ flux at the top of the 
atmosphere 
(hereafter referred to as TOA flux) measured by BESS95 \cite{bess95}. For 
experimental data referring to other measurements with much less statistics see
Refs.\cite{all,hof,mitchell,moiseev,barbiellini}. 
Also displayed in Fig. 1 are the minimal, 
median and maximal fluxes expected for secondary antiprotons at the
time of the BESS95 data taking. These fluxes have been derived 
with a procedure which is described in detail in Secs. II--V. 

A comparison of the BESS95 data with the
theoretically expected fluxes for secondary $\bar p$'s, as displayed in 
Fig. 1, leads to the following 
considerations: i) the experimental data are consistent with the theoretically 
expected secondary flux, within the experimental errors and the theoretical 
uncertainties; however, ii) the experimental flux seems to be suggestive of a
flatter behaviour, as compared to the one expected for secondaries 
${\bar p}$'s. Thus, 
natural questions arise, such as: a) how much room for exotic ${\bar p}$'s 
would there be in the BESS95 data, for instance in case the secondary flux is
approximately given by the median estimate of Fig. 1, b) how consistent 
with the current theoretical models would be
the interpretation of the BESS95 data in terms of a fractional presence of 
exotic antiprotons, and  c) how this interpretation might be checked by means
of independent experiments? 
In the present note we address these questions within an interpretation of a
possible excess of $\bar p$'s at low energies 
in terms of primary antiprotons generated by relic neutralinos in the galactic 
halo \cite{note1}.

The present analysis \cite{note2} is mostly meant to a clarification of many
theoretical points which will be even more crucial, when a much more
statistically significant experimental information on low--energy cosmic--ray
antiprotons will be made available by forthcoming experiments: AMS on the
precursor Shuttle flight in May 1998 and on the International Space Station 
Alpha (ISSA) in January 2002 \cite{ams}, the satellite--borne 
PAMELA experiment \cite{pamela} 
and balloon--borne measurements \cite{spill}.

Our paper is organized as follows. In Sec. II we discuss the cosmic--ray 
IS proton
spectrum which will be subsequently employed in deriving the secondary
antiprotons. In the same section we also illustrate how we treat the solar 
modulation to connect the IS spectra to the corresponding TOA fluxes. In
Sec. III we discuss the sources of cosmic
antiprotons, both of primary and of secondary origins. Cosmic rays
diffusion properties are
derived in Sec. IV; the TOA $\bar p$ spectra are given in Sec. V.
In Sec. VI we compare our theoretical fluxes 
with the BESS95 data and single out the neutralino configurations which may be
relevant for the present problem. Secs. VII and VIII are devoted to an analysis 
on how these supersymmetric configurations  can be explored by direct 
searches for relic neutralinos and by experimental investigation at
accelerators. Conclusions and perspectives in terms of the forthcoming 
measurements of low--energy cosmic--ray $\bar p$'s  are illustrated in Sec. IX.

\section{Cosmic--ray proton spectrum}

We have first to fix the primary IS cosmic--ray proton spectrum, since we need 
it for the evaluation of the secondary $\bar p$'s. 
The IS cosmic--ray proton spectrum is derived by assuming for it appropriate
parametrizations and by fitting their corresponding solar--modulated expressions
to the TOA experimental fluxes. 

Measurements of the TOA spectra have always suffered from large
uncertainties, as discussed for instance in Ref. \cite{gs}. 
In the present paper we use the two most recent high--statistics measurements 
of the TOA proton spectrum:
the one reported  by the IMAX Collaboration on the basis of a balloon flight in 
1992 \cite{imaxp}, the other given by the CAPRICE Collaboration based on data 
collected during a balloon flight in 1994 \cite{caprice}. These two fluxes are
reported in Fig. 2.

For the IS proton spectrum we have used two different parametrizations: 
one in terms of the total proton energy, $E_p = T_p + m_p$,

\begin{equation}
\Phi_p^{\rm IS}(T_p) = A\,\beta (E_p/{\rm GeV})^{- \alpha} \,\,
{\rm \frac{protons}{m^2 \cdot s \cdot sr \cdot GeV}},
\label{eq:energy}
\end{equation}
the other in terms of momentum, $p$ 
(equivalent to rigidity for protons),

\begin{equation}
\Phi_p^{\rm IS}(T_p) = B\,\beta^{-1} (p/{\rm GeV})^{- \gamma} \,\,
{\rm \frac{protons}{m^2 \cdot s \cdot sr \cdot GeV}},
\label{eq:rigidity}
\end{equation}
\noindent
where $\beta = p/E_p$. For the solar modulation effect we
have employed the Perko method \cite{perko}, where the solar--modulated flux 
is given by

\begin{equation}
\Phi^{\rm TOA}(T) = \frac {T^2+2m_pT} {T^2_{\rm IS}+2m_pT_{\rm IS}} 
\Phi^{\rm IS}(T_{\rm IS}).
\label{eq:sm}
\end{equation}

\noindent
The kinetic energies $T$ and $T_{\rm IS}$ are simply related by 
 $ T_{\rm IS} = T + \Delta$, when $ T \geq  T_{\rm cut}$ and by a more 
complicated
relation otherwise \cite{perko}. Thus, this solar--modulation recipe is fully
defined, once the values for the two parameters $\Delta$ and $ T_{\rm cut}$ are 
given.  

The results of our best fits to the data of Refs. \cite{imaxp,caprice} 
are reported in Table I in terms of the parameters of expressions 
 (\ref{eq:energy}) and (\ref{eq:rigidity}) and of the solar--modulation 
parameter $\Delta$. 
The first and third sets of values for parameters $A,
\alpha, \Delta$ (for expression (\ref{eq:energy}))
and $B, \gamma, \Delta$ (for expression (\ref{eq:rigidity})) refer to
3--parameter fits over the entire energy range of the experimental data. 
These fits are mainly meant to fix the solar--modulation parameter $\Delta$,
since the low energy part of the spectra is strongly dependent on the effect of
solar modulation. 
The second and fourth sets of values refer to 2--parameter fits (at fixed 
$\Delta$) in the high energy range  ($T_p \geq 20$ GeV), where the solar
modulation effect is less sizeable, however not negligible, and therefore the
proper parameters of the IS flux (normalization and spectral index) can be
determined more confidently. 

The best--fit values for $T_{\rm cut}$ turn out to be always smaller than the 
value corresponding to the lowest $T$ considered in the fit 
(i.e., $T_{\rm cut} <$ 0.1 GeV). 
This is consistent with the determination of
the cut--off rigidity of the diffusion coefficient in the heliosphere
\cite{perko2,rastoin}. 

From the values reported in Table I we notice that 
even using various parametric forms for the IS proton spectrum, the data
of the two experiments of Refs. \cite{imaxp,caprice} do not lead to a set of
central values for the parameters mutually compatible within their
uncertainties. This difference can be considered as due to systematics in the
measurement of the proton spectra. We also notice that the parametrization of 
Eq.(\ref{eq:rigidity}) systematically provides larger values for the
solar--modulation parameter $\Delta$ as compared to the ones obtained using
the parametrization of Eq.(\ref{eq:energy}). This is due to the steeper
behaviour at low energies of the function of Eq.(\ref{eq:rigidity}) with
respect to Eq.(\ref{eq:energy}).

In Fig. 2 we display the curves of the best fits to the data of 
Refs. \cite{imaxp,caprice} with the parametrization of Eq. (\ref{eq:energy}). 
This is the expression for the IS proton spectrum that will be used hereafter. 
Thus, as a 
median cosmic--ray proton flux we take the expression of Eq.(\ref{eq:energy}),
where the values assigned to the parameters $A$ and $\alpha$ are 
the averages of the central values in the
fits to the data of the two experiments for $T_p \geq 20$ GeV, i.e.

\begin{equation}
\Phi_p^{\rm IS}(T_p) = 15,950\,\beta (E_p/{\rm GeV})^{- 2.76} \,\,
{\rm \frac{protons}{m^2 \cdot s \cdot sr \cdot GeV}}.
\label{eq:mean}
\end{equation}

To estimate the uncertainty in the cosmic--ray proton flux we have combined the
uncertainties of the two parameters $A$ and $\alpha$ 
(at fixed $\Delta$) in our fits at high energies (second set of values in Table
I). We found that a conservative (generous) uncertainty band is delimited by
a minimal flux, given by expression (\ref{eq:energy}) with values: 
$A = 12,300, \alpha = 2.61$, and a maximal flux given by 
 expression (\ref{eq:energy}) with values: 
$A = 19,600, \alpha = 2.89$. Minimal, median and maximal IS proton fluxes are
displayed in Fig. 3 together with the experimental TOA spectra of 
Refs. \cite{imaxp,caprice}. 
It turns out that, for instance, at 100 GeV the uncertainty in the IS proton
flux, relative to the median spectrum, is $\lsim \pm 50 \%$.

\section{Production of antiprotons in the Galaxy}

\subsection{Secondaries $\bar p$'s}

Cosmic ray protons interact with the interstellar material that mostly spreads
in the galactic disk. This conventional spallation is actually a
background to an hypothetical supersymmetric antiproton signal. It needs
therefore to be carefully estimated, especially at low energies where
new measurements are expected. The corresponding source term is
given by the convolution between the antiproton production cross section
and the interstellar proton energy spectrum as
\begin{equation}
\mbox{$q_{\bar{\rm p}}^{\rm disk}$}(r) \; = \;
{\displaystyle \int_{E^{0}_{\rm p}}^{+ \infty}} \,
{\displaystyle \frac{d \sigma_{\rm p H \to \bar{p}}}{d \mbox{$E_{\bar{\rm p}}$}} }
\left\{ \mbox{$E_{\rm p}$} \to \mbox{$E_{\bar{\rm p}}$} \right\} \; n_{\rm H} \; v_{\rm p} \; \mbox{$\psi_{\rm p}$} (r,\mbox{$E_{\rm p}$}) \;
d \mbox{$E_{\rm p}$}
\;\; ,
\label{convolution_secondaries}
\end{equation}
where $n_{\rm H}$ is the hydrogen density in the disk, $v_p$ the proton
velocity and $\mbox{$\psi_{\rm p}$}$ is the proton
density per energy bin at distance $r$ from the galactic centre in the
galactic frame.
The collision takes place between an incoming high energy proton with
an hydrogen atom at rest, lying in the gaseous HI and HII clouds of the
galactic ridge. 
The proton energy is denoted by $\mbox{$E_{\rm p}$}$. It is larger
than the threshold $E^{0}_{\rm p} = 7 \, m$. That spallation
reaction may generate an antiproton with energy $\mbox{$E_{\bar{\rm p}}$}$. The relevant
differential production cross section is the sum over the angle $\theta$
between the incoming proton and the produced antiproton momentum
\begin{equation}
{\displaystyle \frac{d \sigma_{\rm p H \to \bar{p}}}{d 
\mbox{$E_{\bar{\rm p}}$}} }
\left\{ \mbox{$E_{\rm p}$} \to \mbox{$E_{\bar{\rm p}}$} \right\} \; = \;
2 \pi \; P_{\bar {\rm p}} \;
{\displaystyle \int_{0}^{\theta_{\rm max}}} \,
\left. E_{\bar {\rm p}} \, \frac{d^{3} \sigma}{d^{3} P_{\bar {\rm p}}} 
\right|_{\rm LI} \;
d \left( - \cos \theta \right) \;\; ,
\label{integral_production_lab}
\end{equation}
where $P_{\bar {\rm p}} = \sqrt{\mbox{$E_{\bar{\rm p}}$}^{2} - m^{2}}$.
That integral is carried out in the galactic frame, at fixed antiproton
energy $\mbox{$E_{\bar{\rm p}}$}$ . 
The proton energy determines the center of mass frame (CMF) energy
$\sqrt{s} = \left\{ 2 m \, \left( \mbox{$E_{\rm p}$} + m \right) \right\}^{1/2}$.
The latter sets in turn the maximal CMF energy $E^{*}_{\rm {\bar p} \, max}$
which the antiproton may carry away once it is produced
\begin{equation}
E^{*}_{\rm {\bar p} \, max} \; = \;
{\displaystyle \frac{s \, - \, 8 m^{2}}{2 \sqrt{s}} } \;\; .
\end{equation}
The range of angles $\theta$ over which the integral in 
Eq. (\ref{integral_production_lab}) is performed is set by the requirement
that the CMF antiproton energy $E^{*}_{\rm {\bar p}}$ should not exceed
the maximal value $E^{*}_{\rm {\bar p} \, max}$ implied by
kinematics. The Lorentz invariant antiproton production cross section
$E_{\bar {\rm p}} \, {d^{3} \sigma} / {d^{3} P_{\bar {\rm p}}}$
has been parametrized by Tan and Ng \cite{TAN_NG_1} as a function
of the transverse and longitudinal antiproton CMF momenta
$P^{*}_{\bar{\rm p} T}$ and $P^{*}_{\bar{\rm p} L}$. We refer the
interested reader to this analysis. The transverse momentum in the CMF is
equal to $P^{*}_{\bar{\rm p} T} = P_{\bar {\rm p}} \sin \theta$ while
the longitudinal momentum $P^{*}_{\bar{\rm p} L}$ obtains from the
component $P_{\bar {\rm p}} \cos \theta$ after a Lorentz boost from the
galactic frame to the CMF of the reaction.
Note finally that the antiproton production integral in Eq. 
(\ref{convolution_secondaries}) should be a priori performed
everywhere in the confining magnetic fields of the galactic disk. It actually
involves the interstellar proton flux $\mbox{$\Phi_{\rm p}$}$ 
which depends on the location $r$.

\subsection{$\bar p$'s from neutralino annihilation}

The differential rate per unit volume and unit time for the production 
of $ \bar p$'s from $\chi$--$\chi$ annihilation is defined as
\begin{equation}
q_{\bar p}^{\rm susy}(T_{\bar p}) \equiv \frac{d S(T_{\bar p})}{d T_{\bar p}} = 
\, <\sigma_{\rm ann} v> \, g(T_{\bar p})
\left( \frac{ \rho_\chi (r,z)}{m_\chi} \right)^2 ,
\label{eq:source} 
\end{equation}
where $<\sigma_{\rm ann} v>$ denotes the average over the galactic velocity
distribution function of neutralino pair annihilation cross section 
$\sigma_{\rm ann}$ multiplied by the relative velocity $v$ of the annihilating 
particles, $m_\chi$ is the neutralino mass; $g(T_{\bar p})$ denotes 
the $\bar p$ differential spectrum 
\begin{equation}
g(T_{\bar p}) \equiv {1 \over \sigma_{\rm ann}} 
{{d \sigma_{\rm ann} (\chi \chi \rightarrow
 \bar p + X)} \over {dT_{\bar p}}} = \sum_{F,h}
B^{(F)}_{\chi h}{dN^{h}_{\bar p}\over dT_{\bar p}},
\label{eq:propagation}
\end{equation}
where $F$ indicates the $\chi$--$\chi$ annihilation final states,
$B^{(F)}_{\chi h}$ is the branching ratio into
quarks or gluons $h$ in the channel $F$ and
$dN^h_{\bar p}/dT_{\bar p}$ is the differential energy distribution 
of the antiprotons generated by hadronization of quarks and gluons.
In Eq.(\ref{eq:source}), $\rho_\chi (r,z)$ is the mass distribution
function of neutralinos in the galactic halo. Here we consider the
possibility that the halo is spheroidal and we parameterize 
$\rho_\chi (r,z)$ as a function of the radial distance $r$ 
from the galactic center in the galactic plane 
and of the vertical distance $z$ from the galactic plane
\begin{equation} 
\rho_\chi (r,z) = \rho^0_{\chi}\,\, \frac{a^2 + r^2_\odot}{a^2 + r^2 + z^2/f^2},
\label{eq:mass_DF}
\end{equation}
where $a$ is the core radius of the halo, 
$r_\odot$ is the distance of the Sun from the
galactic center and $f$ is a parameter which describes the flattening  of the 
halo.
Here we take the values: $a=3.5$ kpc, $r_\odot = 8$ kpc.
For $f$, which in principle may be in the range $0.1 \leq f \leq 1$, 
 we use the two representative values $f = 0.5, 1$ \cite{olling}. 
The quantity $\rho^0_{\chi}$ denotes the local value of the neutralino
matter density. We factorize it as 
$\rho^0_{\chi} = \xi \rho_l$, where $\rho_l$ is the total local dark 
matter density. Here $\xi$ is evaluated as 
$\xi = {\rm min}(1,\Omega_\chi h^2 / (\Omega h^2)_{\rm min})$, where 
$(\Omega h^2)_{\rm min}$ is a minimal value compatible with observational 
data and with large--scale structure calculations
\cite{bbefms}.  All the results 
of this paper refer to the choice $(\Omega h^2)_{\rm min} = 0.03$. 
The neutralino relic density $\Omega_\chi h^2$ is calculated as a function 
of the
supersymmetric parameters as described in Ref. \cite{omega}. As for the value 
$\rho_l$ of
the total dark matter density, this is calculated by taking into account the
contribution given by the matter density of Eq.(\ref{eq:mass_DF}) to the local
rotational velocity. For instance, in the case of a spherical halo ($f = 1$),
a value of $\rho_l = 0.4$ GeV cm$^{-3}$ is obtained. When $f < 1$ (oblate
spheroidal distribution), $\rho_l$ is given by \cite{bt,ggt}
\begin{equation} 
\rho_l (f) = \rho_l (f = 1)\, \frac{\sqrt{1-f^2}}{f \, 
{\rm Arcsin} \sqrt{1-f^2}}.
\end{equation}

All the quantities depending on the supersymmetric parameters have been
calculated in the framework of the Minimal Supersymmetric extension of the 
Standard Model (MSSM) \cite{susy}, where the
neutralino is defined as the lowest--mass linear superposition of photino 
($\tilde \gamma$), zino ($\tilde Z$) and the two higgsino states
($\tilde H_1^{\circ}$, $\tilde H_2^{\circ}$)
\begin{equation}
\chi \equiv a_1 \tilde \gamma + a_2 \tilde Z + a_3 \tilde H_1^{\circ}
+ a_4 \tilde H_2^{\circ}.
\label{eq:neu}
\end{equation}
For the evaluation of the 
averaged annihilation cross section $<\sigma_{\rm ann} v>$ we have followed the
procedure outlined in Ref. \cite{noi}.
We have considered all the tree--level diagrams which are responsible of
neutralino annihilation and which are relevant to $\bar p$ production,
namely: annihilation into quark--antiquark pairs, into gauge bosons,
into a Higgs boson pair and into a Higgs and a gauge boson. For each final state
we have considered all the relevant Feynman diagrams, which involve the
exchange of Higgs and $Z$ bosons in the s--channel and the exchange
of squarks, neutralinos and charginos in the t-- and u--channels.
Finally, we have included the one--loop diagrams which produce a two--gluon 
final state. For this annihilation channel, we have used the recent results of 
Ref.\cite{bu}. 

The $\bar p$ differential distribution $g(T_{\bar p})$ has been
evaluated as discussed in Ref. \cite{noi}. Here we 
only recall that we have calculated the branching ratios 
$B^{(F)}_{\chi h}$ for all annihilation final states which may produce $\bar
p$'s, dividing these states into two categories: 
i) direct production of quarks and
gluons, ii) generation of quarks through intermediate production of 
Higgs bosons, gauge bosons and $t$ quark. In order to obtain the distributions
$dN^h_{\bar p}/dT_{\bar p}$ the hadronization of quarks and gluons has been
evaluated by using the Monte Carlo code Jetset 7.2 \cite{jet}. 
For the top quark, we have considered it to decay before hadronization.

We summarize now the main features of the MSSM scheme we employ here. 
The MSSM is defined at the electroweak scale as a straightforward 
supersymmetric extension
of the Standard Model. The Higgs sector consists of two Higgs doublets 
$H_1$ and $H_2$, which define two free parameters:
the ratio of the two vacuum expectation values $\tan\beta  \equiv \langle H_2
\rangle/\langle H_1\rangle$ and the mass of one of the three neutral physical
Higgs fields; we choose as a free parameter the mass $m_A$ of the neutral
pseudoscalar Higgs. 
The other parameters of the model are contained in the superpotential, which 
includes all the Yukawa couplings and the Higgs--mixing term $\mu H_1 H_2$,
and in the soft--breaking Lagrangian,  which includes the trilinear and bilinear
breaking parameters and the soft gaugino and scalar mass terms.
In order to deal with a manageable model, 
we impose the following usual relations 
among the parameters at the electroweak scale:
i) all trilinear parameters are set to zero except those of the third family,
which are unified to a common value $A$;
ii) all squarks and sleptons soft--mass parameters are taken as
degenerate: $m_{\tilde l_i} = m_{\tilde q_i} \equiv m_0$;
iii) the gaugino masses are assumed to unify at $M_{GUT}$, and this implies that
the $U(1)$ and $SU(2)$ gaugino masses are related at the electroweak scale by
$M_1= (5/3) \tan^2 \theta_W M_2$.

When all these conditions are imposed, the supersymmetric parameter space is
completely described by six independent parameters, which we choose to
be: $M_2, \mu, \tan\beta, m_A, m_0, A$. In our analyses, we vary them in the
following ranges:
$10\;\mbox{GeV} \leq M_2 \leq  500\;\mbox{GeV}$
(21 steps over a linear grid);
$10\;\mbox{GeV} \leq |\mu| \leq  500\;\mbox{GeV}$
(21 steps, linear grid);
$75\;\mbox{GeV} \leq m_A \leq  500\;\mbox{GeV}$
(15 steps, logarithmic grid);
$100\;\mbox{GeV} \leq m_0 \leq  500\;\mbox{GeV}$
(5 steps, linear grid);
$-3 \leq {\rm A} \leq +3$
(5 steps, linear grid); $1.01 \leq \tan \beta \leq 50$
(15 steps, logarithmic grid).

The supersymmetric parameter space is constrained by 
all the experimental limits obtained from accelerators on supersymmetric and 
Higgs searches. The latest LEP2 data on Higgs, neutralino, chargino and 
sfermion masses  \cite{lep182} and the constraints
due to the $b \rightarrow s + \gamma$ process \cite{alam} are imposed. 
Moreover, the request for the neutralino to be the Lightest 
Supersymmetric Particle (LSP) implies that regions where the gluino or squarks
or sleptons are lighter than the neutralino are excluded. 
A further constraint is imposed by  requiring that all the
supersymmetric configurations  which provide 
a neutralino relic abundance  are in accordance with the cosmological bound
$\Omega_{\chi}h^2 \leq 0.7$.

\section{Diffusion of cosmic rays inside the galaxy}

%
% Propagation equations for the protons. Solution in terms
% of a Bessel expansion.
%
The propagation of cosmic rays inside the Galaxy has been considered
in the framework of a two--zone diffusion model. We have followed here
the same analysis as Webber, Lee and Gupta \cite{webber}. The Milky
Way is pictured as a thin disk, 200 pc across, that extends radially up to
$R = 20$ kpc from the galactic center. That ridge lies between two
extended layers $\sim$ 3 kpc thick, where cosmic rays diffuse in
erratic magnetic fields. Mere diffusion governs the propagation of the
particles in the disk and in the confinement regions that extend on
either side. Assuming that steady state holds, the proton density
$\mbox{$\psi_{\rm p}$}$, per energy bin, at some location $r$ and $z$, is given by
\begin{equation}
\frac{\partial \mbox{$\psi_{\rm p}$}}{\partial t} \; = \; 0 \; = \;
\vec{\nabla} \cdot \left( K \; \vec{\nabla} \, \mbox{$\psi_{\rm p}$} \right) \; + \;
2 h \delta(z) \, q(r) \; - \;
2 h \delta(z) \, \Gamma_{\rm p} \, \mbox{$\psi_{\rm p}$} \;\; .
\label{DIFFUSION_P}
\end{equation}
The diffusion coefficient $K$ is assumed to be essentially independent of
the nature of the species that propagate throughout the Galaxy.
It increases with rigidity $\cal{R}$ according to the relation
\begin{equation}
K({\cal{R}}) \; = \; K_{0} \,
\left( 1 + \frac{\cal{R}}{\cal R}_{0} \right)^{0.6} \;\; ,
\label{DIFFUSION_K}
\end{equation}
where $K_{0} = 6 \times 10^{27}$ cm$^{2}$ s$^{-1}$ and
${\cal R}_{0} = 1$ GV. Below that critical value, the diffusion
coefficient stays constant while above 1 GV, it increases like
${\cal{R}}^{0.6}$. Sources are located in the galactic ridge at
$z = 0$. Their radial profile is inferred from the survey by
Lyne, Manchester and Taylor of the galactic distribution of stellar
remnants and pulsars \cite{LMT} with
$q(r,0) \propto \rho^{\displaystyle a} \, \exp(- b \rho)$ where
$\rho = r/R$, $a = 0.6$ and $b = 3$.
Finally, cosmic ray protons may interact with the interstellar gas.
The latter is assumed to be concentrated in the disk. The probability
per unit time that a proton collides with an interstellar hydrogen
atom at rest is
\begin{equation}
\Gamma_{\rm p} \;=\;
n_{\rm H} \, \sigma_{\rm p H}^{\rm tot} \, v_{\rm p} \;\;.
\end{equation}
The hydrogen density $n_{\rm H}$ is assumed to be constant all over the disk.
The value of $n_{\rm H} = 1$ cm$^{-3}$ is basically consistent with
measurements of the hydrogen column density derived from HI and CO
surveys. It implies in particular a maximal value of
$\sim 9 \times 10^{22}$ H cm$^{-2}$ to be compared to an average
of $5 \times 10^{22}$ H cm$^{-2}$ on the observations of the galactic
center. The densest spot is inferred from CO measurements to reach
a level of $\sim 1.4 \times 10^{23}$ H cm$^{-2}$.
The total interaction cross section $ \sigma_{\rm p H}^{\rm tot}$
between the propagating high energy
protons and the hydrogen atoms of the interstellar medium has been
borrowed from the work by Tan and Ng \cite{TAN_NG_2}.
Above a kinetic energy of 3 GeV, it may be expressed as
\begin{equation}
\sigma_{\rm p H}^{\rm tot} \; = \; (32.2 \; {\rm mb}) \;
\left\{ 1 \; + \; 0.0273 \, U \right\} \;\; ,
\label{SIGMA_PH_TOTAL}
\end{equation}
where the parameter $U$ is defined as
\begin{equation}
U = \ln \left( E_{\rm p} / 200 \, {\rm GeV} \right) \;\; .
\end{equation}
Below $T_{\rm p} = 3$ GeV, expression (\ref{SIGMA_PH_TOTAL}) needs
to be divided by a low energy correction factor equal to
$1 \; + \; 0.00262 \, {T_{\rm p}}^{- C_{\rm p}}$ where
\begin{equation}
C_{\rm p} \; = \; 17.9 + 13.8 \ln T_{\rm p} +
4.41 \ln^{2} T_{\rm p} \;\; .
\end{equation}
The galactic disk is assumed to be infinitely thin, hence the
factor $2 h \delta (z)$ in the diffusion equation (\ref{DIFFUSION_P}),
where $2 h = 200$ pc stands for the actual thickness of the ridge.

Because the Galaxy is axi--symmetric, we can expand the proton
density $\mbox{$\psi_{\rm p}$}$ as a series of Bessel functions of zeroth order
\begin{equation}
\mbox{$\psi_{\rm p}$}(r,z) \; = \;
{\displaystyle \sum_{i = 1}^{+ \infty}} \;
\mbox{${\cal P}_{i}$} (z) \; J_{0} \left( \alpha_{i} \rho \right) \;\;,
\end{equation}
where $\rho = r/R$, while $\alpha_{i}$ is the i--th zero of the
Bessel function $J_{0}$. The proton density is ensured to
vanish at the radial boundary $r = R$ of the system. The Bessel
transforms $\mbox{${\cal P}_{i}$}$ must also drop to
zero at the boundaries of the confinement regions,
at a distance $L = 3$ kpc from either side of the galactic disk.
The distribution of cosmic ray sources may also be expanded
as a series of Bessel functions
\begin{equation}
q(r , E_{\rm p}) \; = \;
{\displaystyle \sum_{i = 1}^{+ \infty}} \;
\frac{q_{i}}{2 \, h} \, Q_{\rm tot}(E_{\rm p}) \;
J_{0} \left( \alpha_{i} \rho \right) \;\; ,
\end{equation}
where $Q_{\rm tot}(E_{\rm p})$ stands for the total galactic
rate of production, per energy bin, of cosmic ray protons with
energy $E_{\rm p}$. The Bessel transforms $q_{i}$ are readily
inferred from the radial distribution of the sources in the galactic
disk
\begin{equation}
q_{i} \; = \;
{\displaystyle \frac{1}{\pi R^{2}} } \;
{\displaystyle \frac{1}{J_{1}^{2} (\alpha_{i})} } \;
\left\{
{\displaystyle \int_{0}^{1} \, q(\rho) \, J_{0}(\alpha_{i} \rho) \, d \rho^{2}}
\right\} \;
\left\{
{\displaystyle \int_{0}^{1} \, q(\rho) \, d \rho^{2}}
\right\}^{-1} \;\; .
\end{equation}

Bessel expanding the diffusion equation (\ref{DIFFUSION_P}) leads to
simple differential relations which the functions $\mbox{${\cal P}_{i}$}(z)$ satisfy.
The latter are even functions of the height $z$ that vanish at the
boundaries of the diffusion layers. A straightforward algebra leads to
\begin{equation}
\mbox{${\cal P}_{i}$} (z , E_{\rm p}) \; = \;
{\displaystyle \frac{q_{i}}{A_{i}}} \; Q_{\rm tot}(E_{\rm p}) \;
{\displaystyle \sinh \left\{ \frac{S_{i}}{2} \left( L - |z| \right) \right\}}
\, / \,
{\displaystyle \sinh \left\{ \frac{S_{i} L}{2} \right\}} \;\; ,
\end{equation}
where $S_{i} = 2 \alpha_{i} / R$ and where the coefficients $A_{i}$ are
defined by
\begin{equation}
A_{i} \; = \; 2 h \Gamma_{\rm p} \; + \;
K S_{i} \, {\displaystyle \coth \left( \frac{S_{i} L}{2} \right)} \;\; .
\end{equation}

Because the diffusion term dominates the behaviour of the
coefficients $A_{i}$, the proton energy spectrum does not vary
much all over the Galaxy, except for a global normalization factor.
In other words, the ratio of the proton fluxes taken at two different
energies is quite insensitive to the location $M$, hence
\begin{equation}
\mbox{$\Phi_{\rm p}$} (M,E_{1}) / \mbox{$\Phi_{\rm p}$} (M,E_{2}) \; \simeq \;
\mbox{${\cal P}_{i}$} (0 , E_{1}) / \mbox{${\cal P}_{i}$} (0 , E_{2}) \;\; .
\end{equation}
This will turn out to be important when we compute the energy spectrum
of secondary antiprotons.

%
% CNO grammage in the galactic disk.
%
The two--zone model is a refinement with respect to the old leaky box
scheme. The confinement layers are necessary in order to account for
the low abundance of the $^{10}$Be unstable element with respect
to its stable partner $^{9}$Be. The former nucleus has a half--lifetime of
1.6 million years (My) and plays the role of a chronometer. Observations
indicate that cosmic rays are trapped in the magnetic fields of our Galaxy for
approximately 100 My before they escape in the intergalactic medium.
On the other hand, the amount of secondary light nuclei such as
lithium, beryllium and boron (Li--Be--B) is well explained by the spallation
of primary carbon, oxygen and nitrogen (CNO) nuclei. The latter spend a mere
5 My in the galactic plane where they cross a column density of
$\sim$ 10 ${\rm g \, cm^{-2}}$. Cosmic rays are therefore confined
most of the time in extended reservoirs above and beneath the matter
ridge, where they just diffuse without interacting much with the
scarce interstellar medium.
We have estimated the grammage which the CNO elements cross
during their journey inside the galactic disk. Their distribution is
inferred in just the same way as for the protons. The average electric
charge per nucleon is now $1/2$ instead of $1$ for the protons, hence a
slightly modified relationship between the kinetic energy per nucleon
and the rigidity of the nucleus under consideration. The grammage is
defined as the product
\begin{equation}
\Lambda_{e} \; = \; n_{\rm H} \, v_{\rm N} \, \tau_{\rm disk} \;\; ,
\end{equation}
where the confinement time in the disk alone is denoted by
$\tau_{\rm disk}$. The escape length $\Lambda_{e}$ is expressed in
units of g cm$^{-2}$. Because cosmic rays either escape from the
disk or interact with its gas, the total number $\mbox{${\cal N}_{\rm N}^{\rm disk}$}$ of particles
contained in the galactic ridge satisfies the balance relation
\begin{equation}
Q_{\rm N} \; = \;
{\displaystyle \frac{\mbox{${\cal N}_{\rm N}^{\rm disk}$}}{\tau_{\rm disk}} } \, + \,
\Gamma_{\rm N} \mbox{${\cal N}_{\rm N}^{\rm disk}$} \;\; .
\end{equation}
The rate $Q_{\rm N}$  at which the CNO primaries are produced
is set equal to the sum of the escape rate from the galactic ridge and of the
interaction rate with the interstellar gas. Notice that in the case of the
two--zone model, the amount $\mbox{${\cal N}_{\rm N}^{\rm disk}$}$ of cosmic rays travelling in the
disk alone may be expressed as the series
\begin{equation}
{\displaystyle \frac{\mbox{${\cal N}_{\rm N}^{\rm disk}$}}{Q_{\rm N}} } \; = \;
4 \pi \, h \, R^{2} \;
{\displaystyle \sum_{i = 1}^{+ \infty}} \;
\frac{q_{i}}{A_{i}} \,
\frac{J_{1} \left( \alpha_{i} \right)}{\alpha_{i}} \;\; .
\end{equation}
In the coefficients $A_{i}$, the relevant cross section that accounts
for the interactions of the CNO species with the interstellar hydrogen
has been averaged at a mere 250 mb. In Fig. 4, an estimate
of the grammage $\Lambda_{e}$ crossed by the CNO elements is presented
as a function of the kinetic energy per nucleon (solid line). It reaches
a maximum of $\sim$ 8 g cm$^{-2}$ at 500 MeV/n. It decreases at
low energies with the velocity. It also drops at high energies
as a result of a better diffusion and hence a lower time of residence in
the disk. The dashed curve refers to the grammage of the protons.
At fixed kinetic energy, the diffusion coefficient is slightly smaller for
these species than for heavier elements, hence a larger escape length
$\Lambda_{e}$.
Measurements of the $^{2}$H abundance have been performed \cite{Seo}
from the Voyager probe at a distance of 23 AU and at energies lying between 20
and 50 MeV/n. With a solar modulation parameter of $\sim$ 360 MV,
this translates into an energy of $\sim$ 230 MeV/n in interstellar space.
The analysis by Seo {\it et al.} of these data is well accounted for by the
leaky box model using a grammage $\Lambda_{e} ({\rm B/C}) \sim$ 8 g cm$^{-2}$.
This is in excellent agreement with the results of our two--zone model
presented in Fig. 4, where the diffusion coefficient $K$
is given by relation (\ref{DIFFUSION_K}).
Ficenec {\it et al.} \cite{Ficenec} have taken data on $^{3}$He between
100 MeV/n and 1.6 GeV/n. They conclude that the grammage of primary
cosmic rays is well fitted by
$\Lambda_{e} = (10.5 + 2.5 - 2.8 ) \; \beta$ g cm$^{-2}$.
The extreme values of that fit are featured by the dotted curves of
Fig. 4. Notice that the CNO grammage inferred
from our two--zone model lies in the range of escape length delineated
by the Ficenec {\it et al.} extreme values, for energies in interstellar space
comprised between 200 MeV/n and 1.5 GeV/n.
The expression which we have adopted for the diffusion coefficient $K$
is therefore well supported by measurements \cite{Ferrando} of the
grammage encountered by primary CNO cosmic rays while they propagate
within the galactic ridge.

%
% Secondary antiprotons.
%
The propagation of antiprotons throughout the Galaxy follows the same
trends as for the protons. We focus first on the species produced by the
spallation of cosmic ray protons with the interstellar gas of the ridge.
Their density \mbox{$\psi_{\bar{\rm p}}$},
per energy bin, follows the diffusion equation
\begin{equation}
\vec{\nabla} \cdot \left( K \; \vec{\nabla} \, \mbox{$\psi_{\bar{\rm p}}$}\right) \; - \;
2 h \delta(z) \, \Gamma_{\bar{\rm p}} \, \mbox{$\psi_{\bar{\rm p}}$}\; + \;
2 h \delta(z) \, \mbox{$q_{\bar{\rm p}}^{\rm disk}$}(r) \; - \;
2 h \delta(z) \, \frac{\partial}{\partial E} \left\{ b(E) \mbox{$\psi_{\bar{\rm p}}$}\right\}
\; = \; 0 \;\; ,
\label{DIFFUSION_PBAR_DISK}
\end{equation}
where steady state has once again been assumed. We recognize the usual
diffusion term as well as the contribution due to the interactions
of the antiproton cosmic rays with the matter of the disk.
The total interaction cross section between antiprotons and the hydrogen
atoms of the interstellar medium has also been borrowed from the analysis
by Tan and Ng \cite{TAN_NG_2}. Above $T_{\bar{\rm p}} = 50$ MeV,
it may be parametrized as
\begin{equation}
\sigma_{\rm \bar{p} H}^{\rm tot} \; = \; (24.7 \; {\rm mb}) \;
\left\{ 1 \, + \, 0.584 \, T_{\bar{\rm p}}^{-0.115} \, + \,
0.856 \, T_{\bar{\rm p}}^{-0.566} \right\} \;\; ,
\label{SIGMA_PBARH_TOTAL}
\end{equation}
where the antiproton kinetic energy $T_{\bar{\rm p}}$ is expressed in GeV.
The spallation source term has already been discussed in Sec. III A.
It obtains from the convolution (\ref{convolution_secondaries}) of the
antiproton production cross section with the proton energy spectrum.
In order to simplify the calculations, we define the effective antiproton
multiplicity
\begin{equation}
N_{\rm \bar{p}}^{\rm eff} (\mbox{$E_{\bar{\rm p}}$}) \; = \;
{\displaystyle \frac{1}{\sigma_{\rm p H}^{\rm tot} (\mbox{$E_{\bar{\rm p}}$})} } \;
{\displaystyle \int_{E^{0}_{\rm p}}^{+ \infty}} \,
{\displaystyle \frac{d \sigma_{\rm p H \to \bar{p}}}{d \mbox{$E_{\bar{\rm p}}$}} }
\left\{ \mbox{$E_{\rm p}$} \to \mbox{$E_{\bar{\rm p}}$} \right\} \;
{\displaystyle \frac{\mbox{$\Phi_{\rm p}$} (\mbox{$E_{\rm p}$})
}{\mbox{$\Phi_{\rm p}$} (\mbox{$E_{\bar{\rm p}}$})} } \; 
d \mbox{$E_{\rm p}$} \;\; .
\end{equation}
Because the ratio of the proton fluxes taken at two different energies does not
depend on the location, the effective antiproton multiplicity
$N_{\rm \bar{p}}^{\rm eff}$ is inferred to be only sensitive to the energy.
It is therefore constant throughout the galactic ridge and may be
computed once and for all as a function of the energy $\mbox{$E_{\bar{\rm p}}$}$ of the
produced antiproton before the diffusion equation (\ref{DIFFUSION_PBAR_DISK})
is solved. The spallation production term readily simplifies into
\begin{equation}
\mbox{$q_{\bar{\rm p}}^{\rm disk}$}(r,E) \; = \; {\sigma_{\rm p H}^{\rm tot} (E)} \;
N_{\rm \bar{p}}^{\rm eff} (E) \; v_{\rm p} \; n_{\rm H} \; \mbox{$\psi_{\rm p}$} (r,E) \;\; .
\end{equation}
Under that form, it may be immediately expanded as the usual series of
Bessel functions of zeroth order.
The last term in relation (\ref{DIFFUSION_PBAR_DISK}) stands for the energy
losses
suffered by the antiproton cosmic rays while they propagate in the galactic
disk. That term actually exists for any cosmic ray species. Because the
particle
fluxes do not significantly drop at low energies, this effect is in general
neglected.
Fluxes tend even to increase below 1 GeV.
In the specific case of secondary antiprotons, that is no longer valid.
Because a high energy proton has very little chance to produce an
antiproton at rest while colliding on an hydrogen atom, the secondary
antiproton flux sharply drops when the energy decreases below $\sim$ 1
GeV. Energy losses tend to shift the antiproton spectrum towards lower
energies with the effect of replenishing the low energy tail with the
more abundant species which had initially a higher energy. This
process is understood here as a mere diffusion in energy space. The
rate at which the antiproton energy varies $b ( \mbox{$E_{\bar{\rm p}}$} ) = \dot{ \mbox{$E_{\bar{\rm p}}$} }$
takes into account two main effects. First, antiprotons may suffer
from ionization losses while they travel across the interstellar gas.
This mechanism yields the following contribution to the energy loss rate
\begin{equation}
b_{\rm \, ion} (E) \; = \; - \, 4 \pi \, r_{e}^{2} \; m_{e} c^{2} \; n_{\rm
H} \;
{\displaystyle \frac{c}{\beta} } \;
\left\{ \ln \left( \frac{2 \, m_{e} c^{2}}{E_{0}} \right) \, + \,
\ln \left( \beta^{2} \gamma^{2} \right) \, - \, \beta^{2} \right\}.
\end{equation}
In molecular hydrogen, the ionization energy
$E_{0}$ has been set equal to 19.2 eV; here
$\gamma = E/m$. The classical radius of the electron is denoted by
$r_{e}$ and the electron mass is $m_{e}$. Second, the dominant
contribution to the energy losses arises from the elastic scatterings of
high energy antiprotons on the hydrogen atoms of the disk. This
mechanism is a counterpart to the collision process whose rate is
$\Gamma_{\bar{\rm p}}$. An antiproton with initial energy $E_{1}$
ends up after such a collision in a final state with the lesser energy
$E_{2}$. Elastic scatterings feed therefore the low energy part of the
antiproton
distribution. They have been described here as if they induced a continuous
change in the antiproton energy. Our assumption is correct on average,
hence the contribution
\begin{equation}
b_{\rm \, scat} (\mbox{$E_{\bar{\rm p}}$}) \; = \; - \,
{\displaystyle \frac{T_{\bar{\rm p}}}{2} } \;
\left\{ \sigma_{\rm \bar{p} H}^{\rm el} (\mbox{$E_{\bar{\rm p}}$}) \, n_{\rm H} \,
v_{\bar{\rm p}} \right\} \;\; .
\end{equation}
The elastic cross section $\sigma_{\rm \bar{p} H}^{\rm el}$
 obtains from the difference
$\sigma_{\rm \bar{p} H}^{\rm tot} \; - \; \sigma_{\rm \bar{p} H}^{\rm an}$
where the annihilation cross section is given by
\begin{equation}
\sigma_{\rm \bar{p} H}^{\rm an} \; = \; (661 \; {\rm mb}) \;
\left\{ 1 \, + \, 0.0115 \, T_{\bar{\rm p}}^{-0.774} \, - \,
0.948\, T_{\bar{\rm p}}^{0.0151} \right\} \;\; ,
\label{SIGMA_PBARH_AN}
\end{equation}
between 100 MeV and 12 GeV, i.e., the energy range under scrutiny
here. Low energy data are fairly consistent with an average energy
loss approximately equal to a half of the initial antiproton kinetic energy
\cite{eisenhandler}.

The antiproton density \mbox{$\psi_{\bar{\rm p}}$}, per energy bin, may be Bessel transformed
into the functions $\mbox{${\bar {\cal P}}_{i}$}$ whose variations with height $z$ are given by
\begin{equation}
\mbox{${\bar {\cal P}}_{i}$} ( \mbox{$E_{\bar{\rm p}}$} , z ) \; = \; \mbox{${\bar {\cal P}}_{i}$} ( \mbox{$E_{\bar{\rm p}}$} , 0 ) \;
{\displaystyle \sinh \left\{ \frac{S_{i}}{2} \left( L - |z| \right) \right\}}
\, / \,
{\displaystyle \sinh \left\{ \frac{S_{i} L}{2} \right\}} \;\; .
\end{equation}
In the galactic disk at $z = 0$, the Bessel transforms $\mbox{${\bar {\cal P}}_{i}$} ( \mbox{$E_{\bar{\rm p}}$} ,
0 )$
only depend on the antiproton energy $\mbox{$E_{\bar{\rm p}}$}$. They actually satisfy a first
order differential equation
\begin{equation}
2 h \, \frac{\partial}{\partial E} \left( b \mbox{${\bar {\cal P}}_{i}$} \right) \; + \;
B_{i} \mbox{${\bar {\cal P}}_{i}$} \; = \; 2 h \, \left. \left(
{\sigma_{\rm p H}^{\rm tot}} \, N_{\rm \bar{p}}^{\rm eff} \, v_{\rm p}
\right) \right|_{E} \; n_{\rm H} \; \mbox{${\cal P}_{i}$} (E , 0) \;\; ,
\label{DIFFERENTIAL}
\end{equation}
which we have numerically solved for each order $i \leq 100$. At high energy,
antiprotons are insensitive to the energy losses. Starting therefore from an
unperturbed spectrum, we have decreased the kinetic energy from 10 GeV
down to 100 MeV while integrating equation (\ref{DIFFERENTIAL}).
The coefficients $B_{i}$ obtain from $A_{i}$ by replacing the rate
$\Gamma_{\rm p}$ by its antiproton counterpart $\Gamma_{\bar{\rm p}}$.
The above mentioned method has been applied to the case of the median
proton flux (\ref{eq:mean}) derived from the IMAX and CAPRICE measurements.
The solid curve of Fig. 5 stands for the corresponding antiproton interstellar
flux. Energy losses have been taken into account. This is not the case however
for the dot--dashed line where the same proton spectrum has been assumed.
Note that energy losses tend actually to replenish the low energy part of the
antiproton distribution. This effect is particularly evident at low energy. For
$\mbox{$T_{\bar{\rm p}}$} \sim$ 100 MeV, the antiproton flux increases by more than an
order of magnitude when energy losses in the gaseous disk are considered.
At larger energies, the upward shift of the spectrum is less sizeable.
For an interstellar kinetic energy of 600 MeV, the increase has reduced to
$\sim$ 30\%.
Even in the case of minimal solar modulation, 600 MeV in interstellar space
translate
into $\mbox{$T_{\bar{\rm p}}$} \sim$ 300 MeV. The dotted and dashed curves 
respectively stand for the antiproton spectra derived from the minimal and 
maximal IS proton flux discussed at the end of Sec.II. 

We finally discuss the case of the antiprotons produced in the annihilation
of neutralinos potentially concealed in the galactic halo. The
diffusion equation is quite similar to relation (\ref{DIFFUSION_PBAR_DISK})
\begin{equation}
\vec{\nabla} \cdot \left( K \; \vec{\nabla} \, \mbox{$\psi_{\bar{\rm p}}$}\right) \; - \;
2 h \delta(z) \, \Gamma_{\bar{\rm p}} \, \mbox{$\psi_{\bar{\rm p}}$}\; + \;
\mbox{$q_{\bar{\rm p}}^{\rm susy}$}(r,z) \; = \; 0 \;\; .
\label{DIFFUSION_PBAR_SUSY}
\end{equation}

Because the energy distribution of these supersymmetric antiprotons
is fairly flat, energy losses in the disk should play a negligible role.
They have not been considered here. The source term (\ref{eq:source})
has already been discussed in Sec. III B. The antiproton production
extends now all over the Galaxy and not solely in the disk. The
solution of the diffusion equation (\ref{DIFFUSION_PBAR_SUSY})
follows however the same trends as for the previous cases. The
antiproton energy distribution \mbox{$\psi_{\bar{\rm p}}$} may still be expanded as a series
of its Bessel transforms $\mbox{${\bar {\cal P}}_{i}$} ( \mbox{$E_{\bar{\rm p}}$} , z )$. Since energy losses are
negligible, the latter obey the simple differential equation
\begin{equation}
K \, \left\{
{\displaystyle \frac{d^{2} \mbox{${\bar {\cal P}}_{i}$}}{dz^{2}} } \, - \,
\frac{\alpha_{i}^{2}}{R^{2}} \, \mbox{${\bar {\cal P}}_{i}$} \right\} \; - \;
2 h \delta(z) \, \Gamma_{\bar{\rm p}} \,  \mbox{${\bar {\cal P}}_{i}$} \; + \;
\mbox{$q_{i}^{\rm susy}$} (z) \; = \; 0 \;\; .
\label{PBAR_SUSY_1}
\end{equation}
The Bessel transforms of the supersymmetric antiproton source
distribution $\mbox{$q_{\bar{\rm p}}^{\rm susy}$}$ are defined as
\begin{equation}
\mbox{$q_{i}^{\rm susy}$} (z) \; = \;
\frac{1}{J_{1}^{2} (\alpha_{i})} \,
{\displaystyle \int_{0}^{1}} \, J_{0}(\alpha_{i} \rho) \,
\mbox{$q_{\bar{\rm p}}^{\rm susy}$}( r = \rho R , z ) \,
d \rho^{2} \;\; .
\end{equation}
Outside the galactic ridge, equation (\ref{PBAR_SUSY_1}) simplifies
even further into
\begin{equation}
{\displaystyle \frac{d^{2} \mbox{${\bar {\cal P}}_{i}$}}{dz^{2}} } \; - \;
\frac{\alpha_{i}^{2}}{R^{2}} \, \mbox{${\bar {\cal P}}_{i}$} \; + \;
\frac{\mbox{$q_{i}^{\rm susy}$} (z)}{K} \; = \; 0 \;\; .
\label{PBAR_SUSY_2}
\end{equation}
The general solution may be expressed as
\begin{eqnarray}
\mbox{${\bar {\cal P}}_{i}$} (z) \; = \; a_{i} \, \cosh \left( \frac{S_{i} z}{2} \right) & + &
b_{i} \, \sinh \left( \frac{S_{i} z}{2} \right) \\ \nonumber
& + & \frac{1}{K S_{i}} \; {\displaystyle \int_{0}^{L}} \,
\exp \left( - S_{i} |z - z'| / 2 \right) \, \mbox{$q_{i}^{\rm susy}$}(z') \, dz' \;\; ,
\end{eqnarray}
where $S_{i} = 2 \alpha_{i} / R$. We leave as an exercise the determination
of the constants of integration $a_{i}$ and $b_{i}$.  They obtain from the
requirement that the Bessel transforms $\mbox{${\bar {\cal P}}_{i}$}$ vanish at the boundaries
$z = \pm L$ of the confinements regions that extend on either side of the
ridge.
Because the antiproton distribution is an even function of the height $z$,
the interested reader may also show that
$\dot{\mbox{${\bar {\cal P}}_{i}$}} (0) \; = \; h \Gamma_{\bar{\rm p}} \mbox{${\bar {\cal P}}_{i}$}(0) / K$.
The final result readily obtains as
\begin{equation}
\mbox{${\bar {\cal P}}_{i}$} (z) \; = \; \frac{2}{K S_{i}} \;
\left\{ {\cal F}(L) \, \frac{{\cal G}(z)}{{\cal G}(L)} \; - \; {\cal F}(z)
\right\}
\;\; .
\end{equation}
This expression describes the actual propagation of antiprotons which
have been produced in remote regions of the halo and that propagate
backwards in the magnetic fields of the Galaxy. The functions ${\cal F}(z)$ and
${\cal G}(z)$ are respectively defined by
\begin{equation}
{\cal F}(z) \; = \; {\displaystyle \int_{0}^{\displaystyle z}} \,
\sinh \left( \frac{S_{i}}{2} |z - z'| \right) \; \mbox{$q_{i}^{\rm susy}$} (z') \; dz'
\;\; ,
\end{equation}
and
\begin{equation}
{\cal G}(z) \; = \;
2 h \Gamma_{\bar{\rm p}} \, \sinh \left( \frac{S_{i} z}{2} \right)
\; + \;
K S_{i} \, \cosh \left( \frac{S_{i} z}{2} \right) \;\; .
\end{equation}
The interstellar flux at the solar system of the antiprotons produced
by the annihilation of hypothetical supersymmetric species comprising
part of the galactic halo may now be expressed as
\begin{equation}
\mbox{$\Phi_{\bar{\rm p}}$} ( \odot , \mbox{$T_{\bar{\rm p}}$} ) \; = \;
<\sigma_{\rm ann} v> \, g(\mbox{$T_{\bar{\rm p}}$}) \,
\left\{ \frac{ \rho_{0}}{m_{\chi}} \right\}^2 \;
C_{\rm susy} ( \mbox{$T_{\bar{\rm p}}$} , f ) \;\; .
\label{SUSY_ANALYSIS}
\end{equation}
The density of reference $\rho_{0}$ has been set equal to 1 GeV cm$^{-3}$.
The coefficient $C_{\rm susy} ( \mbox{$T_{\bar{\rm p}}$} , f )$ is defined as
\begin{equation}
C_{\rm susy} ( \mbox{$T_{\bar{\rm p}}$} , f ) \; = \;
\frac{1}{4 \pi} \, v_{\bar{\rm p}} \, \mbox{$\psi_{\bar{\rm p}}$}^{\rm eff} ( \odot , \mbox{$T_{\bar{\rm p}}$} )
\;\; .
\end{equation}
The effective energy distribution 
$\mbox{$\psi_{\bar{\rm p}}$}^{\rm eff}$ is taken at
the solar circle and has been derived with the above mentioned
method where an effective antiproton source term
$\left\{ \rho_{\chi} (r,z) / \rho_{0} \right\}^{2}$
has been assumed. The latter term depends on the flattening $f$ of
the halo. Note that $C_{\rm susy}$ is not a flux of particles. It is a
mere coefficient that is actually expressed in units of cm sr$^{-1}$.
Fig. 6 illustrates the behaviour of this coefficient when the antiproton
kinetic energy is varied from 100 MeV up to 10 GeV, for three
different values of the flattening factor $f =$ 0.1, 0.5 and 1. The
coefficient $C_{\rm susy} ( \mbox{$T_{\bar{\rm p}}$} , f )$ exhibits a smooth maximum
around $\mbox{$T_{\bar{\rm p}}$} \sim$ 1 GeV. Below that value, it tends to decrease
with the antiproton velocity like $v_{\bar{\rm p}} / B_{i}$.
For higher energies, the diffusion takes place more efficiently and
the cosmic rays escape more easily from the galactic magnetic fields, hence a
lower density in the disk. When the flattening increases, the dark
matter halo is compressed towards the ridge. There are many more
neutralinos in the diffusion layers where antiprotons are kept confined,
hence a larger flux.
The evaluation of the IS ${\bar p}$ flux due to neutralino annihilation is then
performed by using Eq.(\ref{SUSY_ANALYSIS}).

\section{Primary and secondary antiproton TOA fluxes}

Our TOA antiproton fluxes are derived from the corresponding IS spectra, by
employing the Perko solar--modulation procedure \cite{perko}, already defined in
Sec. II. In that section we also derived the values for the parameter $\Delta$
relevant to the  measurements of Refs.\cite{imaxp,caprice}. 
In order to obtain the
$\Delta$ values to be applied in case of experiments performed at different 
times, we use the results of Papini, Grimani and Stephens 
(PGS) \cite{papini}. 
These authors derived  simple analytic expressions as best fits to the
measured  spectra of the TOA primary cosmic--ray protons, 
obtained from a 
large collection of data over a couple of solar cycles.
They provide the parameters of these fits for 
periods of maximum and minimum solar activity. By fitting their analytic
expressions with the solar--modulated flux derived from 
our parametric form of Eq.(\ref{eq:energy}),
we find the
following average values for $\Delta$ at minima and maxima: 
$\Delta_{\rm min} = 320$ MeV and $\Delta_{\rm max} = 800$ MeV, respectively.

In Fig. 7 we plot the time variation of the solar--modulation parameter
$\Delta$, as obtained by our best fit to the experimental data. The 
full circles
represent the best--fit values to the PGS average fluxes at minima (MIN) and at
maxima (MAX) and to the fluxes of Refs.\cite{imaxp,caprice}. 
The open circle refers to the BESS95 data taking period. 
The cross denotes the extrapolated 
value at the time relevant for the AMS measurements with the Shuttle flight.

In Figs. 8 and 9 we display how the effects of the flux distorsion at low
energies, induced by solar modulation, is much stronger for the primary flux 
than for the secondary  one. This may be simply understood in terms of the 
nature
of Eq.(\ref{eq:sm}) and of the different shapes for secondaries and primaries.

\section{Comparison with experimental data}

Early measurements of cosmic--ray antiprotons have been plagued
by low--statistics problems and brought to serious conflicting results at
low energies ($T_{\bar p} \lsim 0.4$ GeV) in the past \cite{hist}. 
As already mentioned in our introduction, a recent analysis \cite{bess95} 
of the data collected by the BESS spectrometer during its 1995 flight   
(BESS95) has provided a significant improvement in statistics in the low--energy
region, with a total of 43 $\bar p$'s in the kinetic--energy range 
$180 ~ {\rm MeV} \leq T_{\bar{p}} \leq 1.4 ~{\rm GeV}$ \cite{note}. This then
allows an interpretation of the experimental data in terms of 
theoretical models in a more meaningful way than in the past. A further 
substantial breakthrough in this direction will be provided by the forthcoming
measurements with AMS \cite{ams}, the satellite--borne 
PAMELA experiment \cite{pamela} and balloon--borne measurements \cite{spill}.

The BESS95 data are displayed in Fig. 1 and compared with our theoretical
evaluations for secondary antiprotons. Our curves are derived according to the
procedure outlined in previous sections. Solar modulation is evaluated at the
time of the BESS95 measurement. 
The band delimited by the dotted and the dashed curves provides the uncertainty
in the secondary $\bar p$'s flux due to the corresponding uncertainty in the
primary IS cosmic ray proton flux (see Sec. II). It turns out that this
uncertainty is $\leq \pm 30\%$ for $T_{\bar p} \leq 2 $ GeV and it grows up to 
$\pm 50\%$ at  $T_{\bar p} \simeq$ 10 GeV. 

From a first look at Fig. 1 it is apparent that the experimental data are
consistent with the flux due to secondary $\bar p$'s. This is indeed
quantitatively confirmed by a $\chi^2$--evaluation, which shows that our median
curve for secondaries fits the BESS95 data with a low reduced--$\chi^2$ value:
 $(\chi^2)_{\rm red}$ = 0.83 (for 5 d.o.f.) \cite{chi2}.

However, it is interesting to explore which would be the chances for a signal, 
due to relic neutralino annihilations, 
of showing  up in the low--energy window ($T_{\bar p} \lsim$ 1 GeV). 
This point is very
challenging, especially in view of the interplay which might occur among
low--energy 
measurements of cosmic--ray $\bar p$'s and other searches, of quite a different 
nature, for relic neutralinos in our Galaxy. 

Actually, 
we find that the agreement between BESS95 experimental data and 
theory may be improved by adding a fraction of neutralino--induced  $\bar p$'s
to the standard secondary antiprotons. 
The best fit to the experimental data with a total theoretical flux 
$\Phi^{\rm th} = \Phi^{\rm sec}_{\rm med}$ + $\Phi^{\rm susy}$, performed by 
varying the
supersymmetric parameters over the grid defined in Sec. IIB,  
provides a value $(\chi^2)_{\rm red}$ = 0.28, with an improvement over the 
$(\chi^2)_{\rm red}$ previously obtained by using the secondary flux only.
This fact cannot certainly be taken as significant of an evidence 
of a neutralino--induced antiproton signal, but shows that indeed the 
low--energy
region $\bar p$ spectrum is still a quite interesting window for exploring
$\bar p$'s of supersymmetric origin, and encourages further investigation of the
problem. 

Now we wish to specifically determine which regions of the 
supersymmetric  parameter space
 (and then which neutralino configurations) might be relevant for 
the problem at hand and how these could be investigated by other experimental
means.   As 
a quantitative criterion to select the relevant supersymmetric configurations, 
we choose to pick up only the configurations 
which meet the following requirements: i) they generate a total theoretical
flux $\Phi^{\rm th}$ which is at least at the level of the experimental value 
(within 1-$\sigma$) in the first energy bin; ii) their 
$(\chi^2)_{\rm red}$, in the best fit of the BESS95 data, is bounded by 
$(\chi^2)_{\rm red} \leq$ 2.2 (corresponding to 95\% C. L. for 5 d.o.f.). 
This set of configurations is hereafter denoted as set $M$; its subset, whose
$\Omega_{\chi} h^2$ values fall in the cosmologically interesting range 
$ 0.03 \leq \Omega_{\chi} h^2 \leq 0.7$, is denoted as set $N$.
An example of a fit to the BESS95 
data which includes a neutralino--induced signal
 with a $(\chi^2)_{\rm red} \leq$ 2.2 is shown in Fig. 10. 
This signal corresponds to a neutralino with the following properties: 
$m_{\chi} = 62$ GeV, $P = 0.98$ and $\Omega_{\chi} h^2 = 0.11$.   

On the other hand, supersymmetric configurations with a 
$(\chi^2)_{\rm red} >$ 4 have to be considered strongly disfavoured by BESS95 
data (actually, they are excluded at 99.9 \% C.L.). We call $R$ 
this set of supersymmetric configurations and we will discuss them  later on.
Supersymmetric configurations belonging neither to $M$ nor to $R$ can only
provide a $\bar p$ flux fully buried in the secondary $\bar p$ flux and are 
then completely irrelevant for the problem under discussion. 

The composition of configurations in sets $M$ and $N$ are displayed in 
Fig. 11 (Fig. 12), where $\tan \beta$ ($m_{\chi}$) is plotted in terms 
of the fractional amount of 
gaugino fields, $P = a_1^2 + a_2^2$, in the neutralino mass eigenstate. 
From Fig. 11 we notice that gaugino configurations are 
more numerous than others, with only a slight  correlation with 
$\tan \beta$; the
requirement of a sizeable contribution to the relic density introduces a 
noticeable reduction in the number of higgsino--like and mixed 
configurations. Fig. 12 shows that 
higgsino--like and mixed configurations are much stronger constrained in the
neutralino mass range than the gaugino--like ones. 
In Fig. 13 we display the features of configurations of set $R$. These
configurations, which are to be considered excluded on the basis of the 
BESS95 data, turn out to be gaugino--like with masses on the low side. 

Up to now, we have discussed our results in terms of a spherically symmetric
galactic halo. The effect of a flattening in the dark matter distribution is to
enhance the primary $\bar p$ flux. Since the size of this flux is proportional
to the function $C_{\rm susy}(T_{\bar p},f)$, defined in Sec. IV, 
the enhancement of the primary
flux as a function of $f$ may be read directly from Fig. 6. For instance, for
$f$ = 0.5 the enhancement factor is 2.3. This has consequences on the nature of
configurations in sets $M, N$ and $R$. By way of example, we plot in Fig. 14
the scatter plot for configurations of set $R$ for a flattening of 
$f$ = 0.5. This
may be compared with the corresponding plots of Fig. 13 which refer to $f$ = 1.
Obviously, the enhancement of the primary flux, induced by the halo flattening,
increases the number of excluded configurations. 

In the present paper we have considered only uniform dark matter distribution 
inside the density profile of Eq. (\ref{eq:mass_DF}). As is well known, any
effect of local density enhancement or clumpiness would induce a substantial
increase in the primary $\bar p$ spectrum, as in any other signal due to pair
annihilation taking place in the halo \cite{clumpy}.

Let us now examine whether our relevant neutralino configurations may be
explored in terms of direct detection experiments for particle dark matter
candidates.

\section{Exploration by direct detection of relic particles}

The measurements of the energy differential rates in experiments of direct 
search for particle dark matter enable the 
extraction of an upper bound for the neutralino--nucleon scalar cross--section 
$\sigma_{\rm scalar}^{\rm nucleon}$, multiplied
by the neutralino local (solar neighbourhood) density, i.e.
 an upper bound for 
the quantity  $\xi \sigma_{\rm scalar}^{\rm nucleon}$, 
once a specific value to the 
total local dark matter density is assigned \cite{dic}. 
By combining all
present experimental data \cite{cald}, we obtain the (90 \% C.L.) upper bound 
displayed in  Fig. 15 by the open solid
curve (the total local dark matter density is normalized
here and in the rest of this paper to the value $\rho_l = 0.4$ GeV cm$^{-3}$). 
The experiments which are
essential in the determination of this upper bounds, in the neutralino mass
range considered here, are those of 
Refs. \cite{few}. The region in Fig. 15 delimited by a closed contour is the
one singled out by the experiment of Ref. \cite{dama} as possibly indicative of
an annual modulation effect (for an interpretation of these data 
in terms of relic neutralinos
see Ref. \cite{bdfs}). 
The scatter plot displays the values of $\xi \sigma_{\rm scalar}^{\rm nucleon}$
 for the configurations of set $M$ (part (a) of Fig. 15) and of set
$N$ (part (b)). It is most remarkable that a sizeable fraction of the
configurations are accessible to investigation by direct detection, since 
the sensitivity in this kind of experiments is expected to be significantly
improved in the near future \cite{cald}. The dashed line in Fig. 15 shows
the discovery potential in case of an improvement by a factor of 10 in current
sensitivities, what is within reach in a short time.   Our analysis shows an 
interesting interplay between experiments
of direct search for particle dark matter and measurements of low--energy $\bar
p$'s in space. This property would obviously be dramatically reinforced, should
the indication about a possible annual modulation effect be confirmed by new
data. In fact, it is very intriguing that many configurations
of set $M$ are indeed in the region singled out by the experiment of Ref.
\cite{dama}. Finally, we notice that some configurations are actually 
excluded
by the direct--search upper bound. This put 
emphasis on the potentiality of direct
detection measurements in providing information on dark matter searches of 
different nature. 

Part (b) 
of Fig. 15 shows how the requirement of a sizeable contribution to the relic
abundance makes somewhat thinner the set of configurations contributing to 
the highest values  of 
$\xi \sigma_{\rm scalar}^{\rm nucleon}$, but still leaves a significant 
number of configurations 
inside the closed region and, anyway, close to the current upper bound curve. 
Correlation between $\xi \sigma_{\rm scalar}^{\rm nucleon}$  and the 
neutralino relic density is given in Fig. 16.

\section{Search at accelerators}

Let us turn now to the question of whether configurations of sets $M$ and $N$
might be explored at accelerators. 

LEP at $\sqrt{s} = 192$ GeV may explore the configurations with a neutralino
mass up to $\simeq 50$ GeV \cite{alt}. Then,
from Fig. 12 we see that LEP will be able to investigate only
marginally the configurations of set $M$ and $N$
in the gaugino sector. Experimental investigation of larger
masses requires future upgrading of the Tevatron or LHC. For instance, 
TeV33 could, under favorable hypothesis, explore a range up to 
$m_{\chi} \simeq 125$ GeV \cite{tev}. In this case, all the higgsino
configurations can be explored, as well as a large portion of the gaugino
sector. 

A further illustrative point is offered by a scatter plot of set $M$ in the
plane $m_{h}$--$\tan \beta$, displayed in Fig. 17 ($m_h$ is the mass of the
lightest CP--even scalar Higgs boson). The representative points 
of the set cover almost completely the Higgs physical region. Part of these 
supersymmetric configurations 
(the ones on the left side of the solid curves) will be explored by LEP at 
$\sqrt{s} = 192$ GeV and at $\sqrt{s} = 200$ GeV, with a luminosity of 
200 $\rm pb^{-1}$ per experiment \cite{alt}.

\section{Conclusions}

We have presented a new  analysis of the cosmic--ray antiprotons flux, expected
on the basis of secondary $\bar p$'s, generated by interactions of cosmic--ray 
primaries with the interstellar medium, 
and of a possible exotic primary source of
$\bar p$'s, originated by neutralino--neutralino annihilations in the Galactic
halo. 

Improvements over previous calculations of secondaries depend mostly on:
i) the use of a two--zone propagation model for diffusion of
cosmic rays in the halo instead of the standard leaky box model;
ii) the inclusion of an energy--loss effect in the propagation properties of
cosmic rays (important for the antiproton low energy range considered in this
paper); 
iii) the use of the new data on primary cosmic--ray proton spectrum, as 
measured by IMAX  \cite{imaxp} and CAPRICE \cite{caprice}. 

The neutralino--induced $\bar p$ flux has been evaluated in a MSSM at the
electroweak scale, which incorporates all current accelerator constraints. 
 Use of supergravity--inspired unification conditions  at large energy
scale has been avoided in order not to arbitrarily constrain the neutralino
phenomenology \cite{bere}. 

Solar modulation of the antiproton flux has been improved by analyzing the most
complete set of data over the solar cycles \cite{papini} and the data on the 
proton spectrum of Refs. \cite{imaxp,caprice}.

We have found that the most statistically relevant data on cosmic--ray
antiprotons at low--energy \cite{bess95} leave some room for a possible signal
from neutralino annihilation in the galactic halo. We have discussed how the
relevant supersymmetric configurations may be explored with direct experiments 
for particle dark matter search and at accelerators. We have shown how the
interplay between measurements of cosmic--ray $\bar p$'s and direct search
experiments for relic particles is very intriguing and quite important in view
of the significant improvements expected in these two classes of experiments in
the near future.
The present analysis stresses the
great interest for the forthcoming AMS measurements with the Shuttle flight
and on the ISSA \cite{ams}, as well as for other future measurements with 
balloon--borne experiments (IMAX\cite{mitchell}, BESS\cite{moiseev})
and with satellites (PAMELA) 
\cite{pamela}, for disentangling the secondary $\bar p$ flux from a
possible primary signal of exotic nature. As an example, we give in Fig. 18 
the distribution of measurements expected for AMS with the Shuttle flight 
according to
two different hypothesis: a) dominance of the secondary contribution
(lower sequence  of crosses), b) significant contribution due a
neutralino--induced signal (upper sequence of crosses).
In our evaluation of the expected measurements we have taken into account 
geomagnetic cut--off effects and the expected AMS overall acceptance
\cite{batt}.

\begin{center}
{\bf Acknowledgments}
\end{center}

We thank Roberto Battiston and Aldo Morselli 
for  interesting discussions about  experimental aspects related to the present
paper. P.S. would like to thank the French Programme National de
Cosmologie for its support.

\newpage
\begin{table}
\caption{Values of the parameters in the expressions (\ref{eq:energy}) and 
(\ref{eq:rigidity}) for the IS proton flux and of the solar--modulation 
parameter $\Delta$. These values are obtained by best--fitting the
data of Refs.[19-20] with 
Eqs. (\ref{eq:energy}) and (\ref{eq:rigidity}), either over the entire energy
range or only over the high--energy ($T_p \geq 20$ GeV) range. First and third sets
of values refer to 3--parameters fits (with 
Eqs.(\ref{eq:energy}) and (\ref{eq:rigidity}), respectively), second and fourth
sets refer to 2--parameters fits at fixed $\Delta$, (with 
Eqs.(\ref{eq:energy}) and (\ref{eq:rigidity}), respectively).
}
\begin{center}
\begin{tabular}{|c|c|c|c|}   \hline

           &     IMAX          &    CAPRICE     &  Comments \\ \hline \hline

 A         & 12,300$\pm$1,700  & 17,600$\pm$500 & entire energy\\ 
 $\alpha$  & 2.67$\pm$0.03     & 2.81$\pm$0.01  &   range    \\
 $\Delta$  & 510$\pm$40        & 390$\pm$ 5     &            \\ \hline
  
 A         & 12,300$\pm$3,000  & 19,600$\pm$ 3,000 &    \\
 $\alpha$  & 2.67$\pm$0.06     & 2.85$\pm$0.04     & $T_p \geq$ 20 GeV  \\
 $\Delta$  & 510 (fixed)       & 390 (fixed)       &        \\ \hline

 B         & 16,200$\pm$2,000  & 26,000$\pm$1,200 & entire energy\\ 
 $\gamma$  & 2.73$\pm$0.03     & 2.91$\pm$0.02    &   range    \\
 $\Delta$  & 795$\pm$35        & 710$\pm$10       &         \\ \hline
  
 B         & 13,700$\pm$4,100  & 22,800$\pm$ 3,700 &      \\
 $\gamma$  & 2.69$\pm$0.06     & 2.88$\pm$0.04     &  $T_p \geq$ 20 GeV  \\
 $\Delta$  & 795 (fixed)       & 710 (fixed)       &     \\ \hline
\end{tabular}
\end{center}
\end{table}

\vfill
\eject

\newpage

\begin{center}
{\bf Figure Captions}
\end{center}

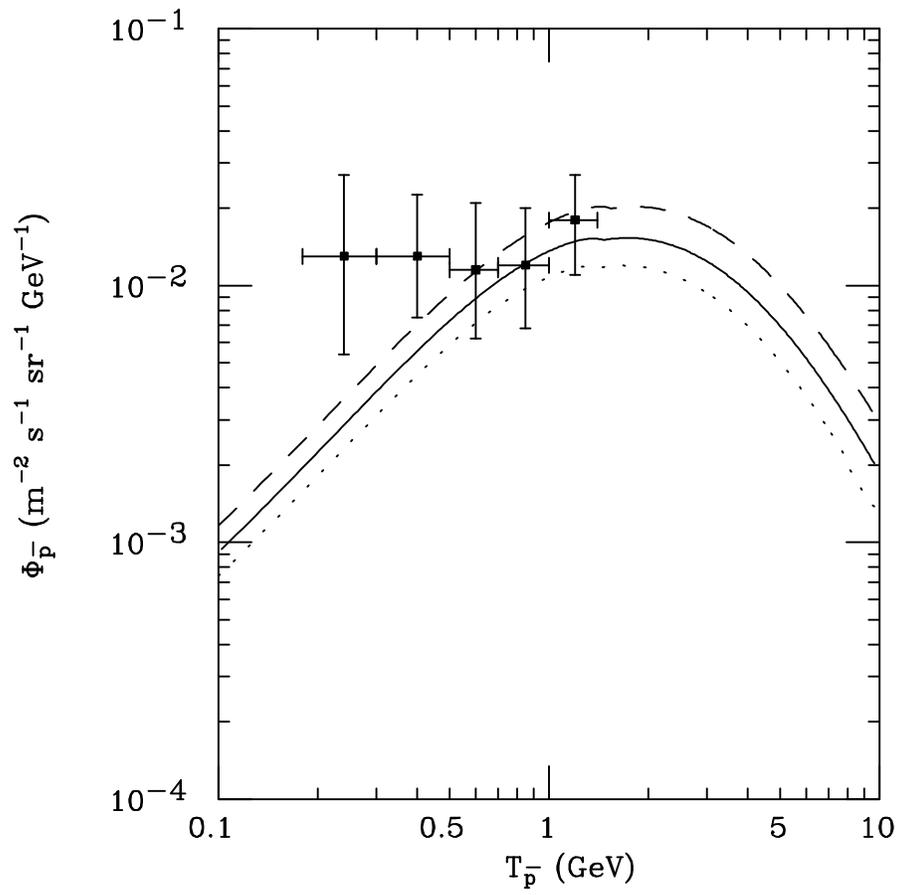
{\bf Fig. 1} - TOA antiproton flux as a function of the antiproton kinetic
energy. The experimental points are the BESS95 data \cite{bess95}. The curves
are the median (solid line), minimal (dotted line) and maximal (dashed line)
secondary fluxes calculated in this paper, solar--modulated at the time of the
BESS95 measurement. 

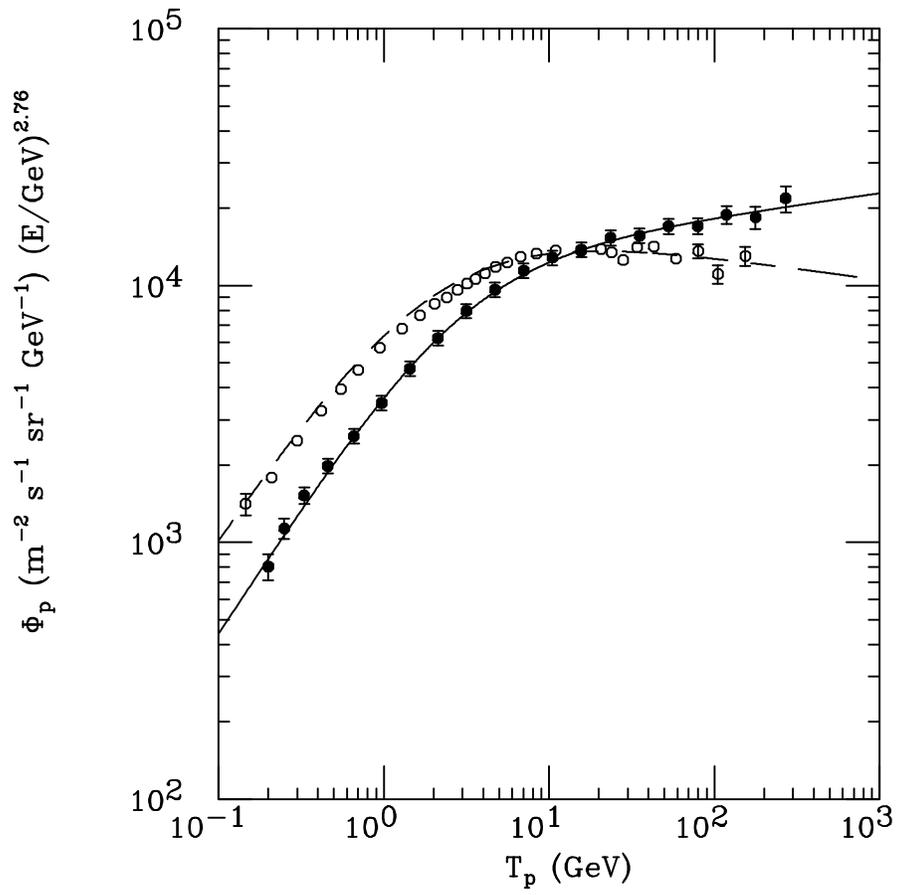
{\bf Fig. 2} - TOA spectra of IMAX (full circles) \cite{imaxp} and of 
CAPRICE (open circles) \cite{caprice} with our best--fit curves with
parametrization of Eq. (\ref{eq:energy}). (The error bars are not shown when 
they are smaller than the dimension of the circles.)

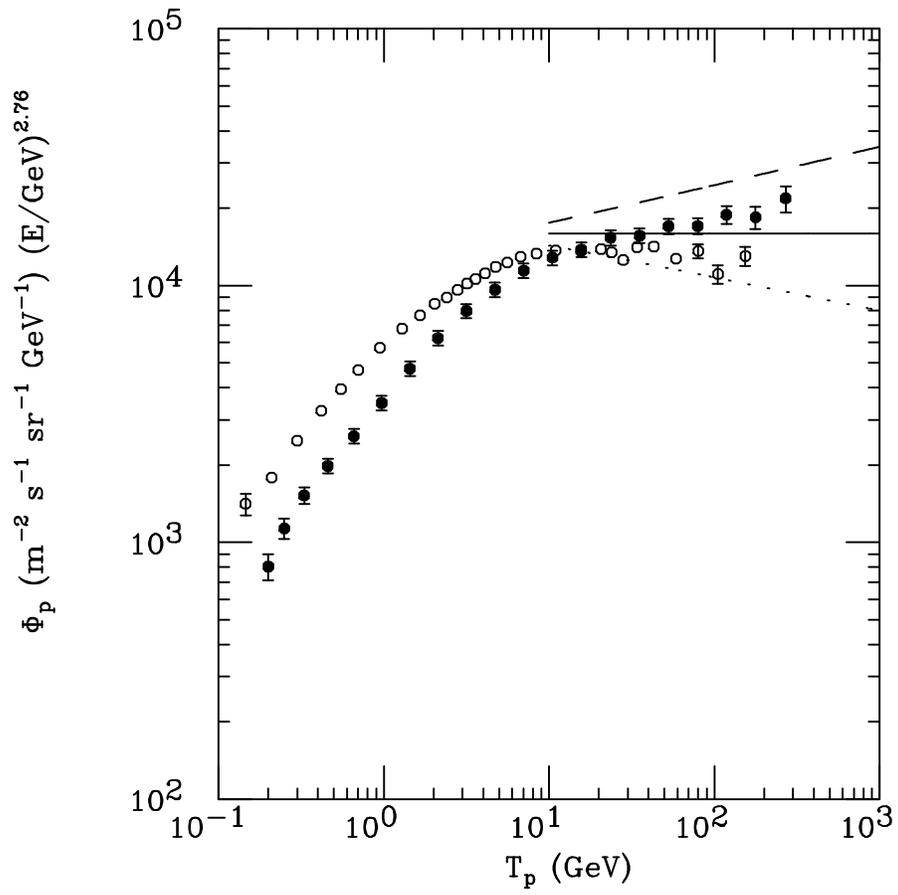
{\bf Fig. 3} - TOA spectra of IMAX (full circles) \cite{imaxp} and of 
CAPRICE (open circles) \cite{caprice}. The solid, dotted and dashed lines denote
the median, minimal and maximal IS proton fluxes, respectively, as discussed in
Sec. II. (The error bars are not shown when
they are smaller than the dimension of the circles.)

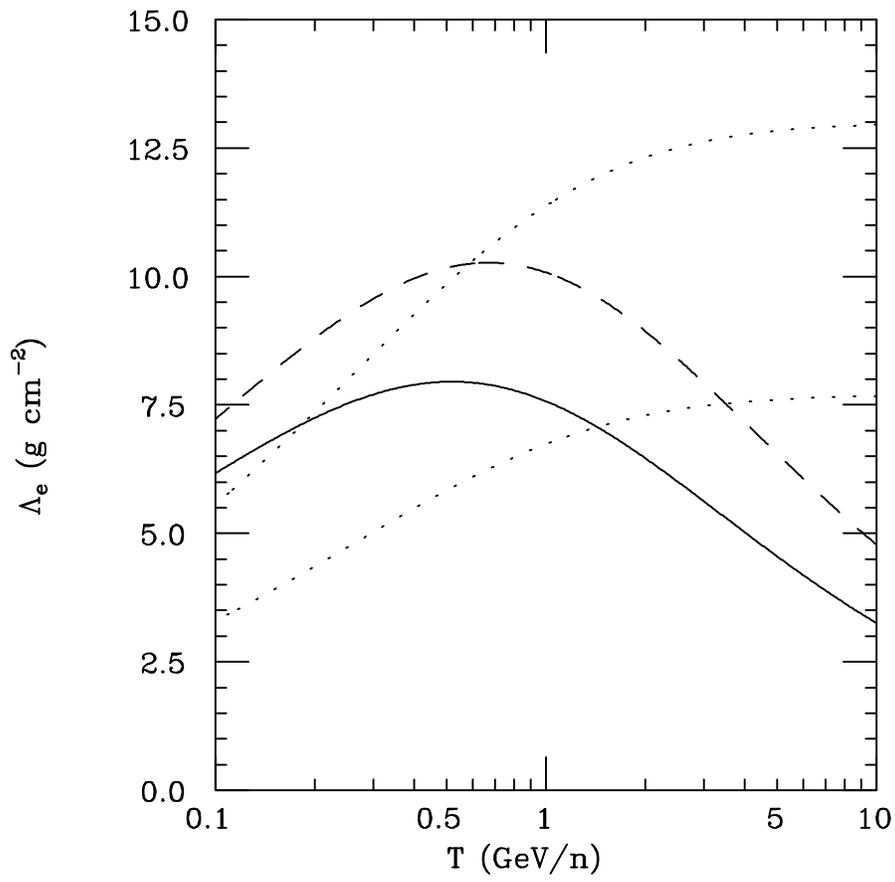
{\bf Fig. 4} - The grammage $\Lambda_e$
of the CNO primary elements (solid) as inferred
from a two--zone diffusion model of the propagation of cosmic rays in
the Galaxy. It is plotted as a function of the kinetic energy per nucleon.
The dashed curve features the grammage corresponding to protons
while the dotted lines delineate the interval of escape lengths inferred from
the Ficenec {\it et al.} \cite{Ficenec}
 observations on $^{3}$He at TOA energies
comprised between 100 MeV/n and 1.6 GeV/n.
\label{fig:grammage}

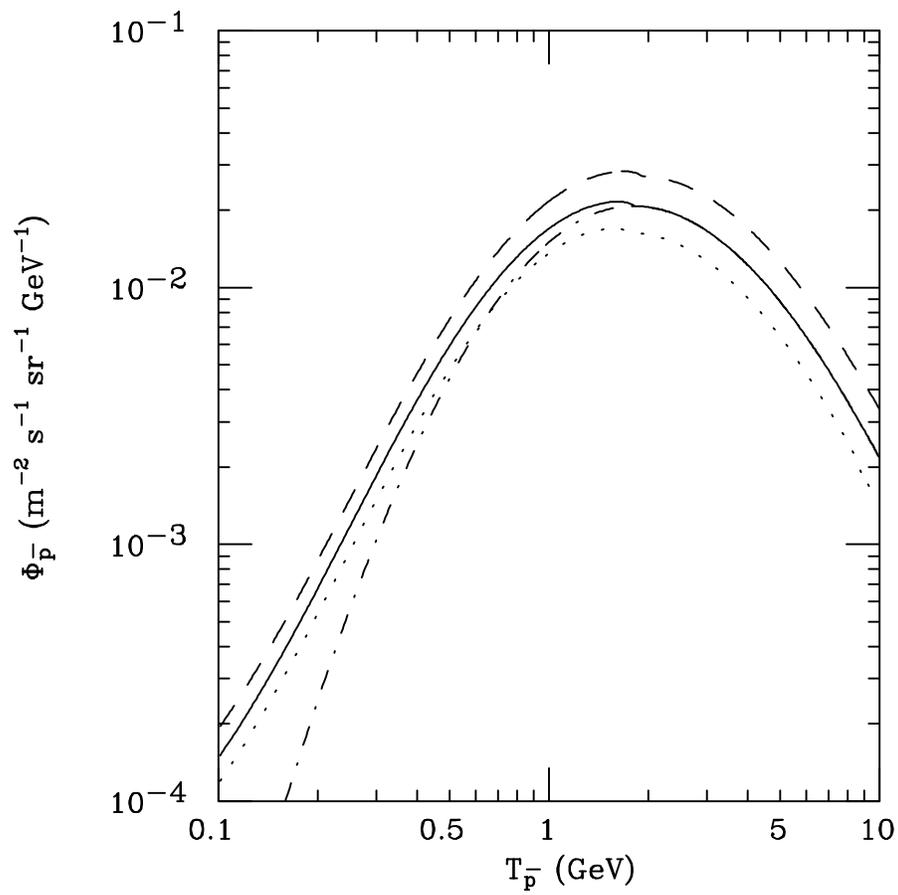
{\bf Fig. 5} - IS secondary antiproton spectra as functions of the ${\bar p}$
kinetic energy. Solid, dotted and dashed lines denote the fluxes obtained from
the median, minimal and maximal IS primary proton fluxes. The dot--dashed line
denotes the median ${\bar p}$ spectrum, when the ${\bar p}$ energy losses are
neglected.  

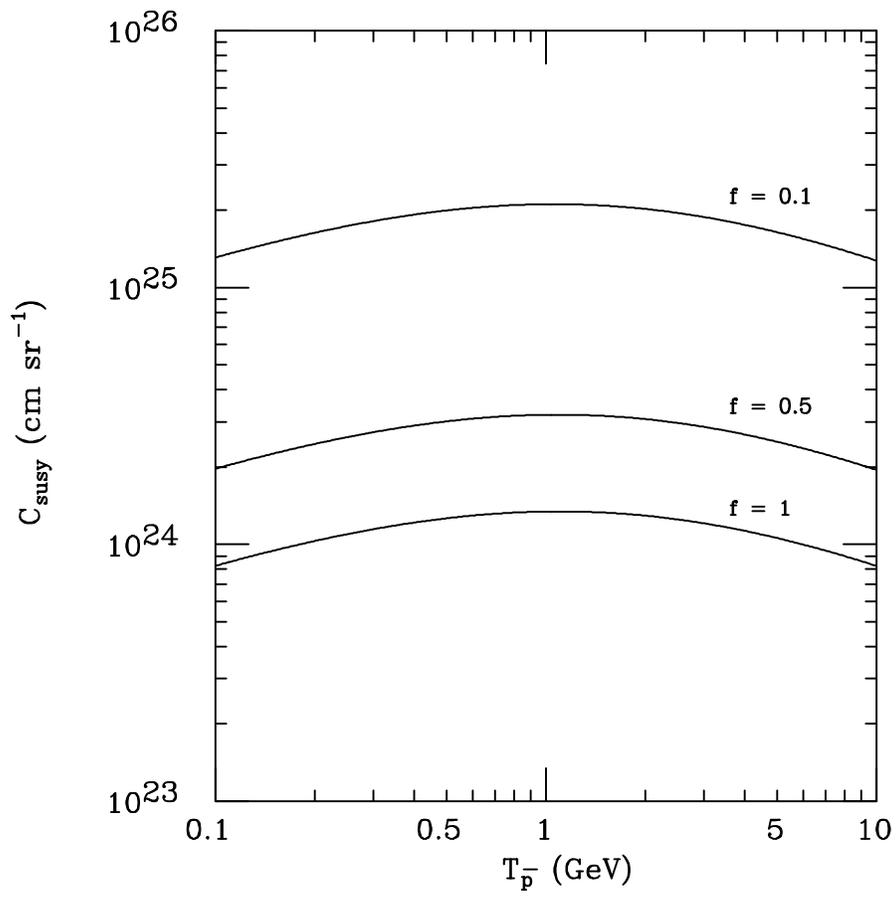
{\bf Fig. 6} - Coefficient $C_{\rm susy}(T_{\bar p},f)$ as a function of the
${\bar p}$ kinetic energy for different values of the flattening parameter $f$. 
 
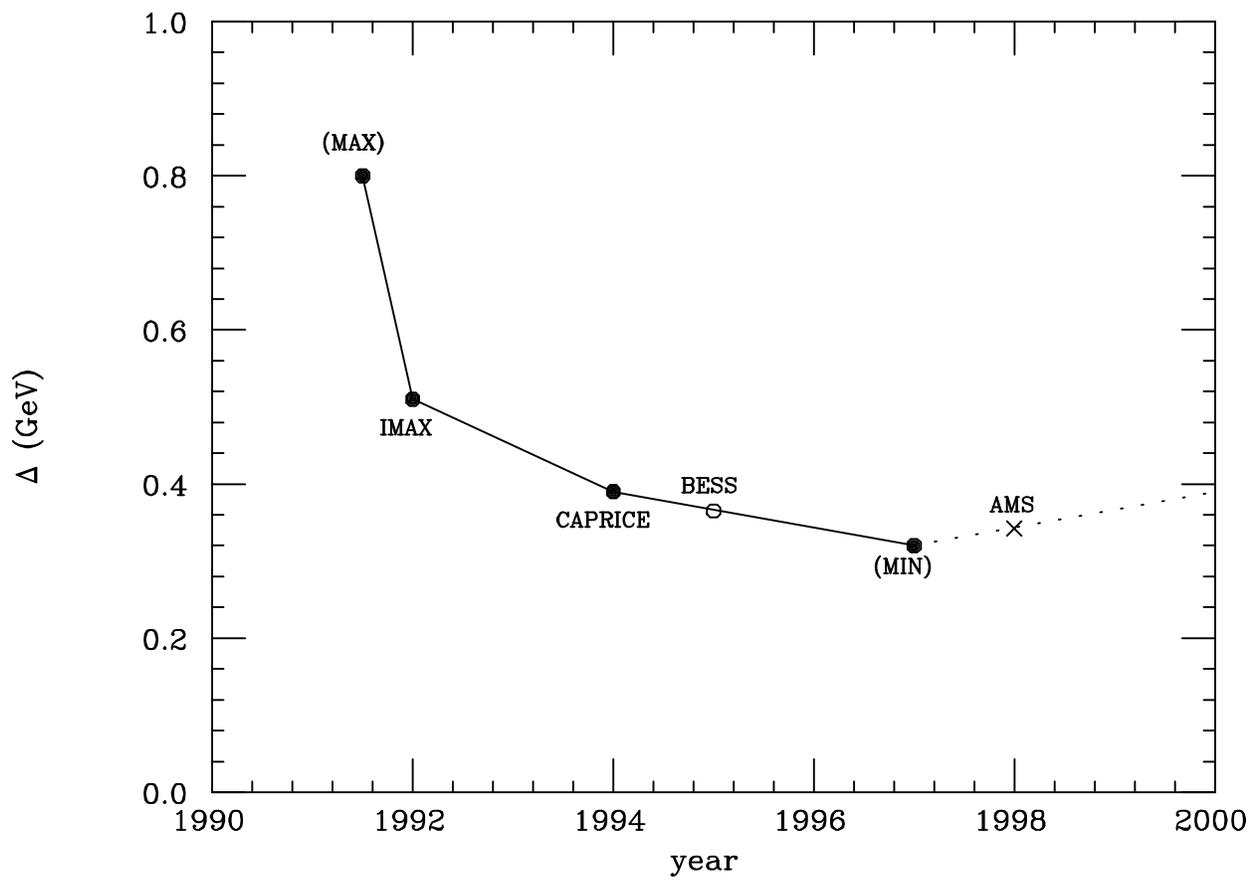
{\bf Fig. 7} - Time variation of the solar--modulation parameter $\Delta$. 
Full circles 
represent the best--fit values to the PGS average fluxes at minima (MIN) and at
maxima (MAX) and to the fluxes  of IMAX \cite{imaxp} and of CAPRICE 
\cite{caprice}; the open circle refers to the BESS95 data taking period and the 
cross denotes the extrapolated  value of $\Delta$ 
at the time relevant for the future AMS Shuttle flight (May 1998). 

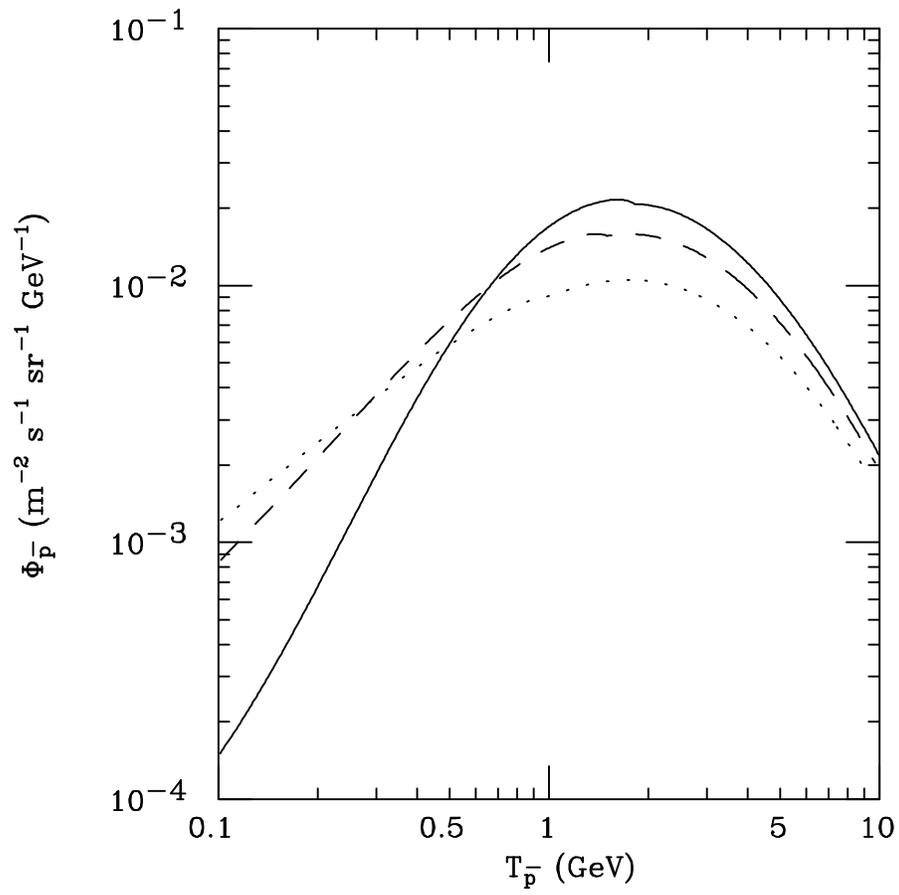
{\bf Fig. 8} - Solar modulation of the IS median secondary antiproton flux
calculated in this paper. Solid line is the IS spectrum; dashed (dotted) 
line is the solar--modulated spectrum at minima (maxima).
 
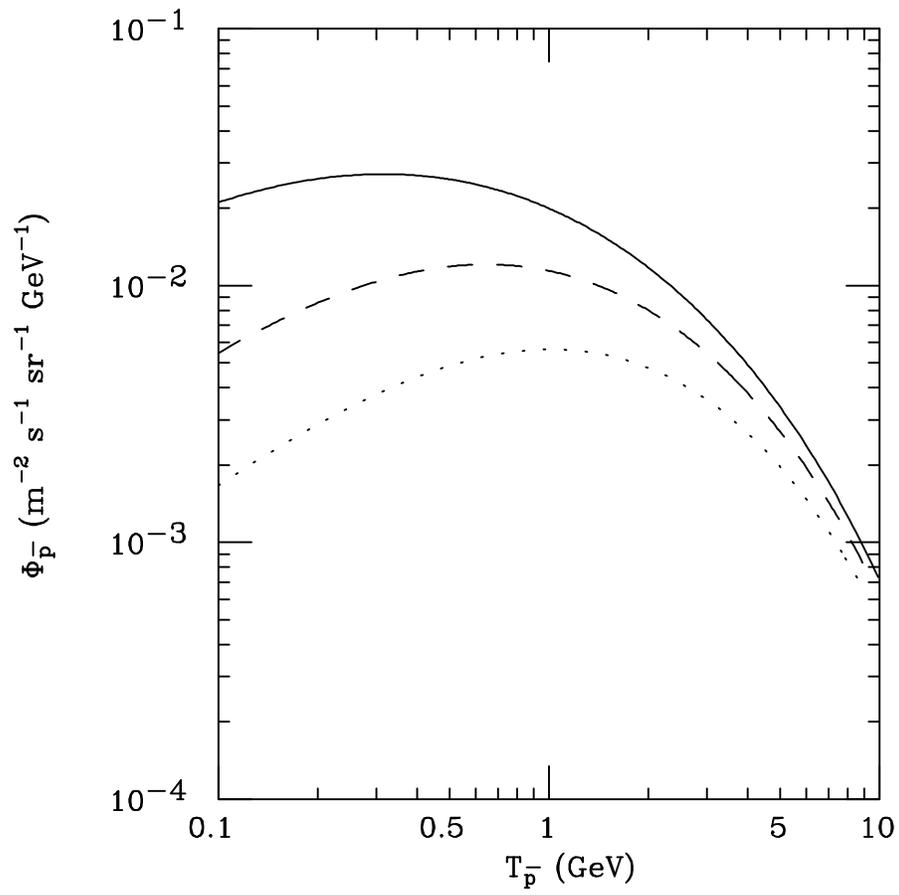
{\bf Fig. 9} - Solar modulation of the IS antiproton flux, due to neutralino
annihilation for the representative neutralino configuration with 
$m_{\chi} = 62$ GeV, $P = 0.98$ and $\Omega_{\chi} h^2 = 0.11$.
Solid line is the IS spectrum; dashed (dotted) 
line is the solar--modulated spectrum at minima (maxima).

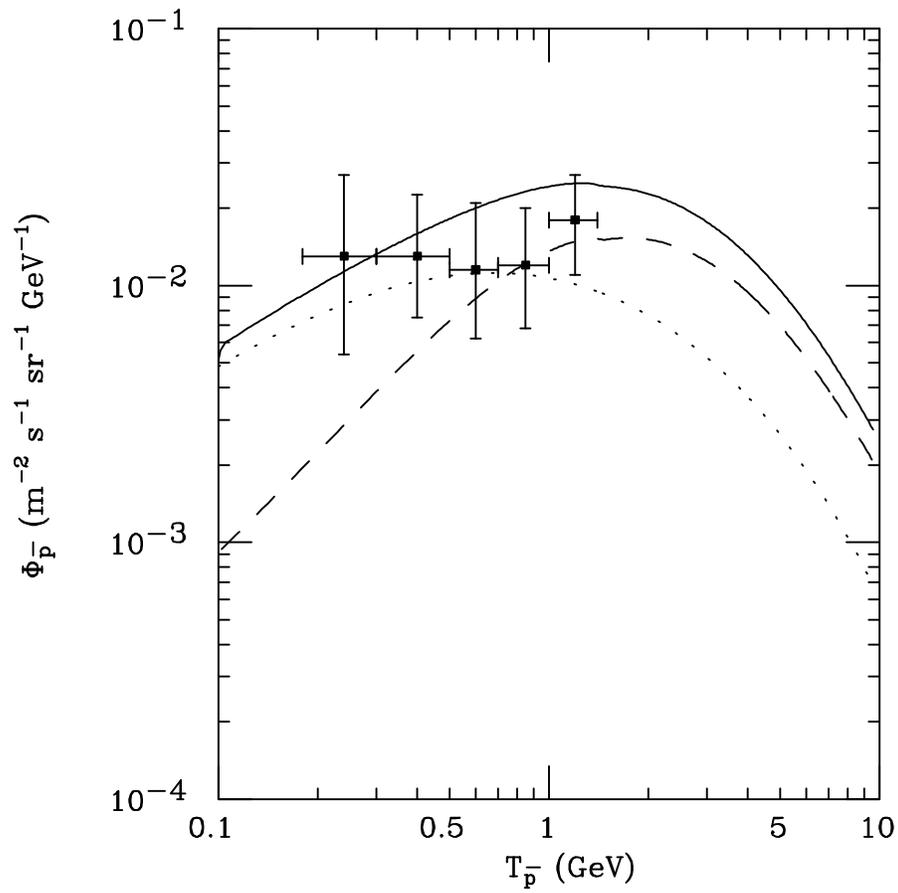
{\bf Fig. 10} - TOA antiproton fluxes versus the antiproton kinetic energy. 
The BESS95 data \cite{bess95} are shown by crosses. 
The dashed line denotes the median secondary flux, the dotted one 
denotes the primary flux due to neutralino annihilation in the halo for a
neutralino configuration with 
$m_{\chi} = 62$ GeV, $P = 0.98$ and $\Omega_{\chi} h^2 = 0.11$.
Solid line denotes the calculated total flux. 

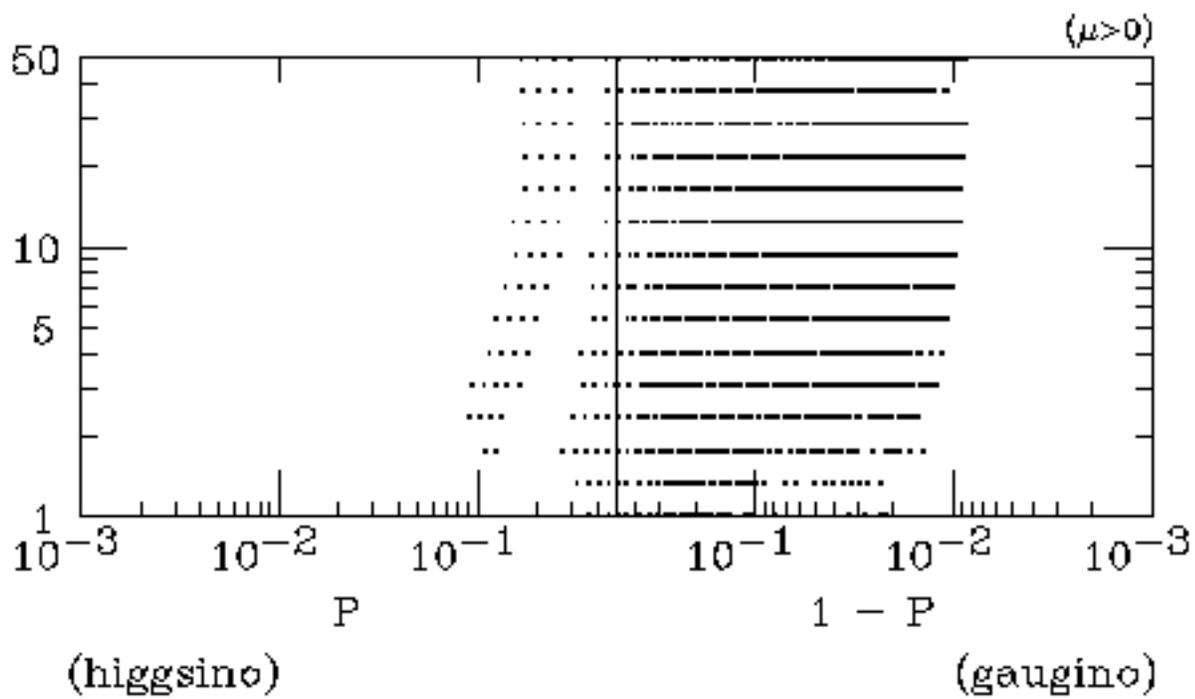
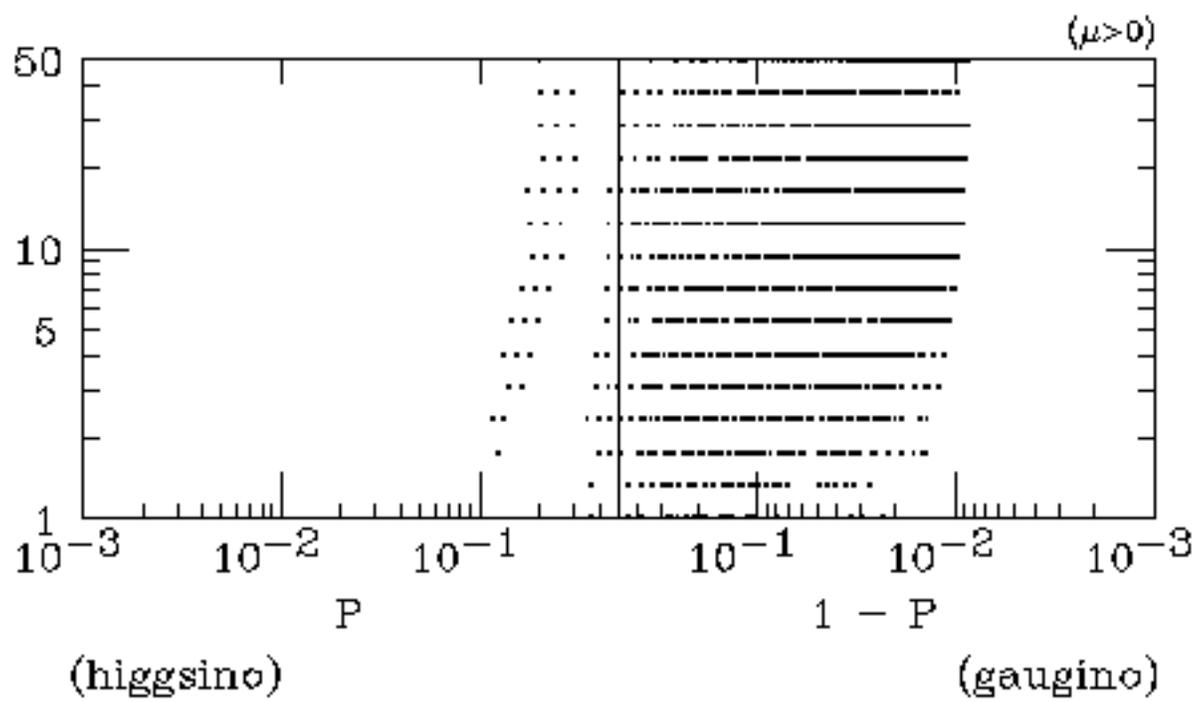
{\bf Fig. 11} - Scatter plots for configurations of set $M$ (a) and
set $N$ (b) in the P--$\tan \beta$ plane.

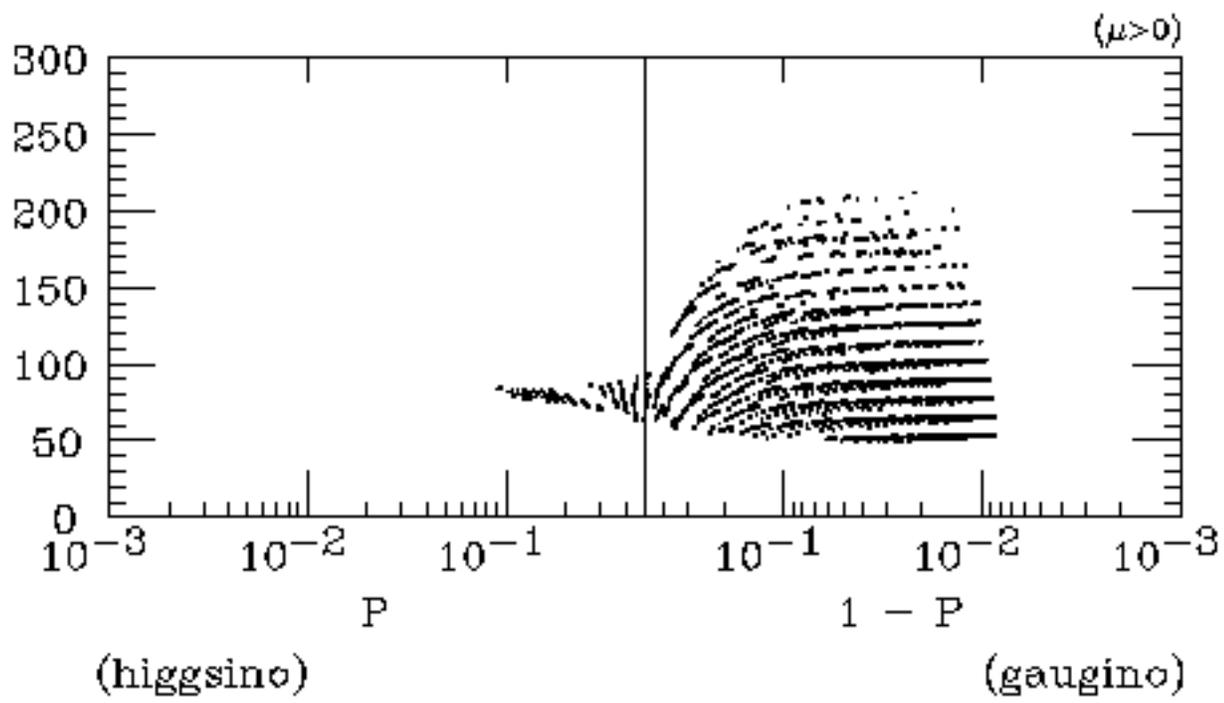
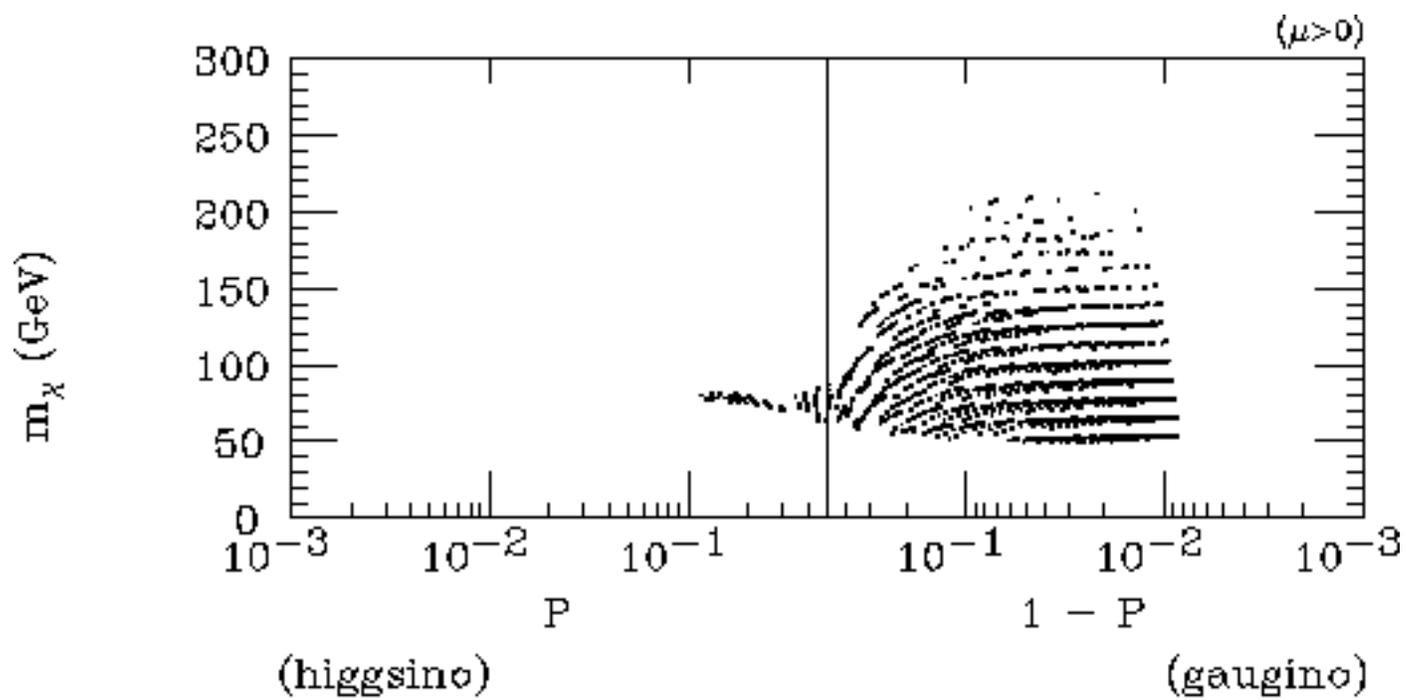
{\bf Fig. 12} - Scatter plots for configurations of set $M$ (a) and
set $N$ (b) in the P--$m_{\chi}$ plane.

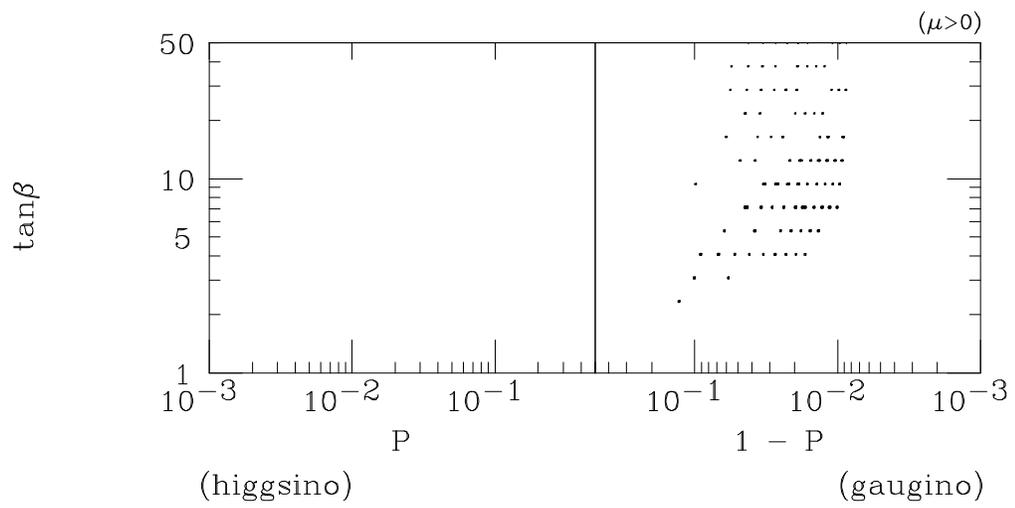
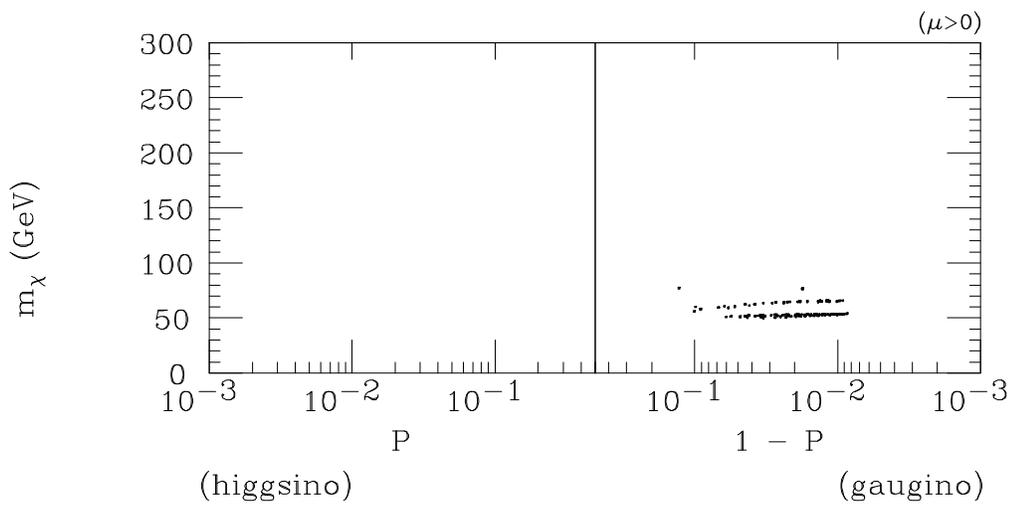
{\bf Fig. 13} - Scatter plots for configurations of set $R$ in the 
 P--$\tan \beta$ plane (a) and in the P--$m_{\chi}$ plane 
(b).

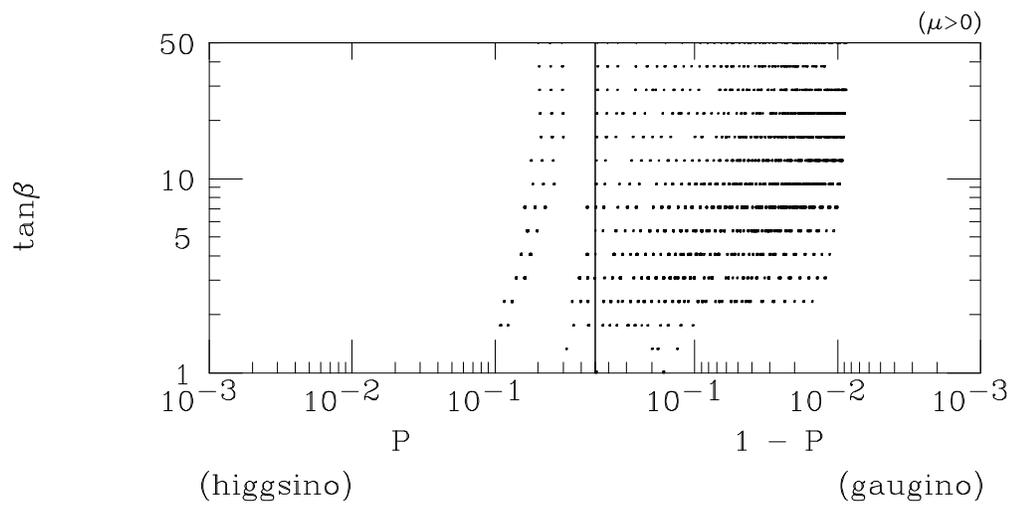
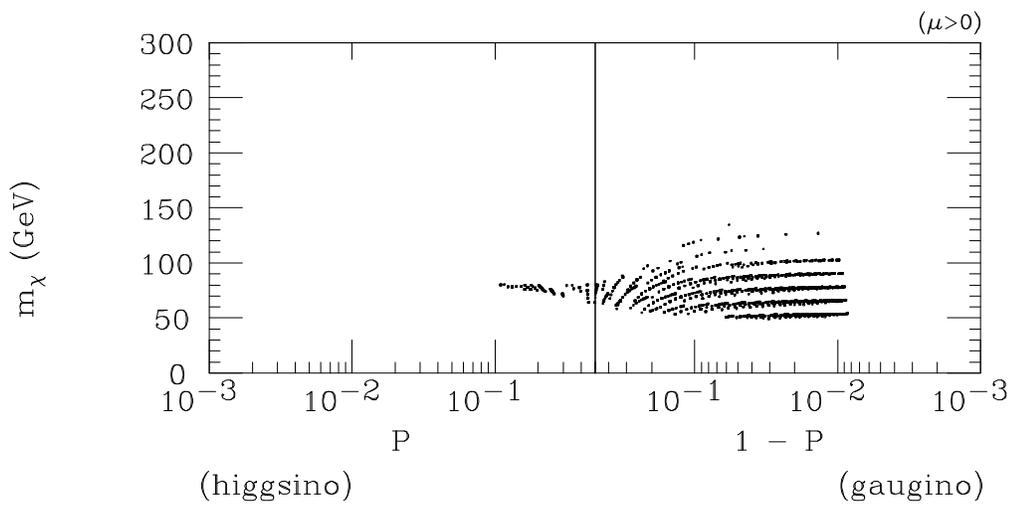
{\bf Fig. 14} - Scatter plots for configurations of set $R$ in the 
 P--$\tan \beta$ plane (a) and in the P--$m_{\chi}$ plane 
(b) for a flattening of $f = 0.5$. 

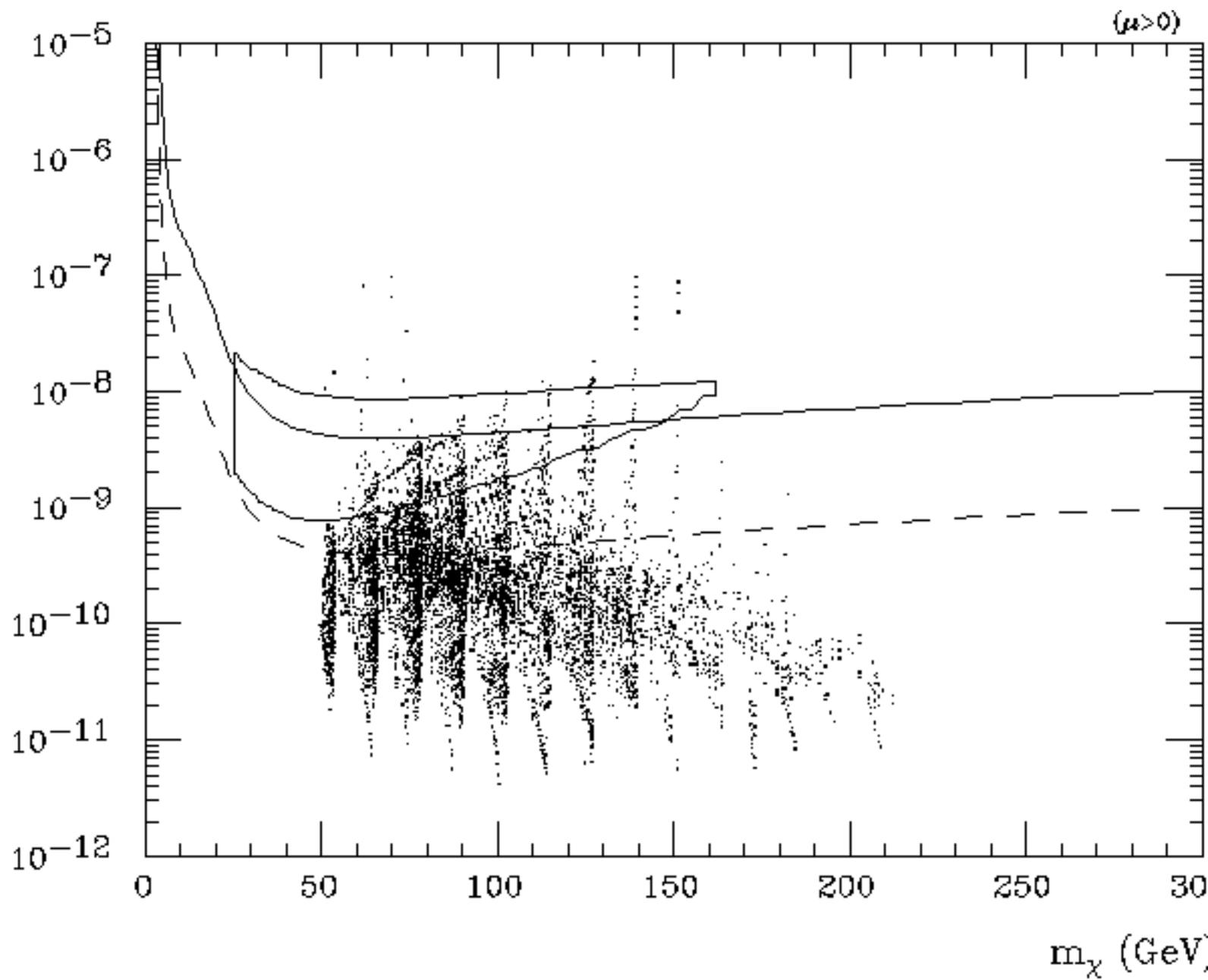
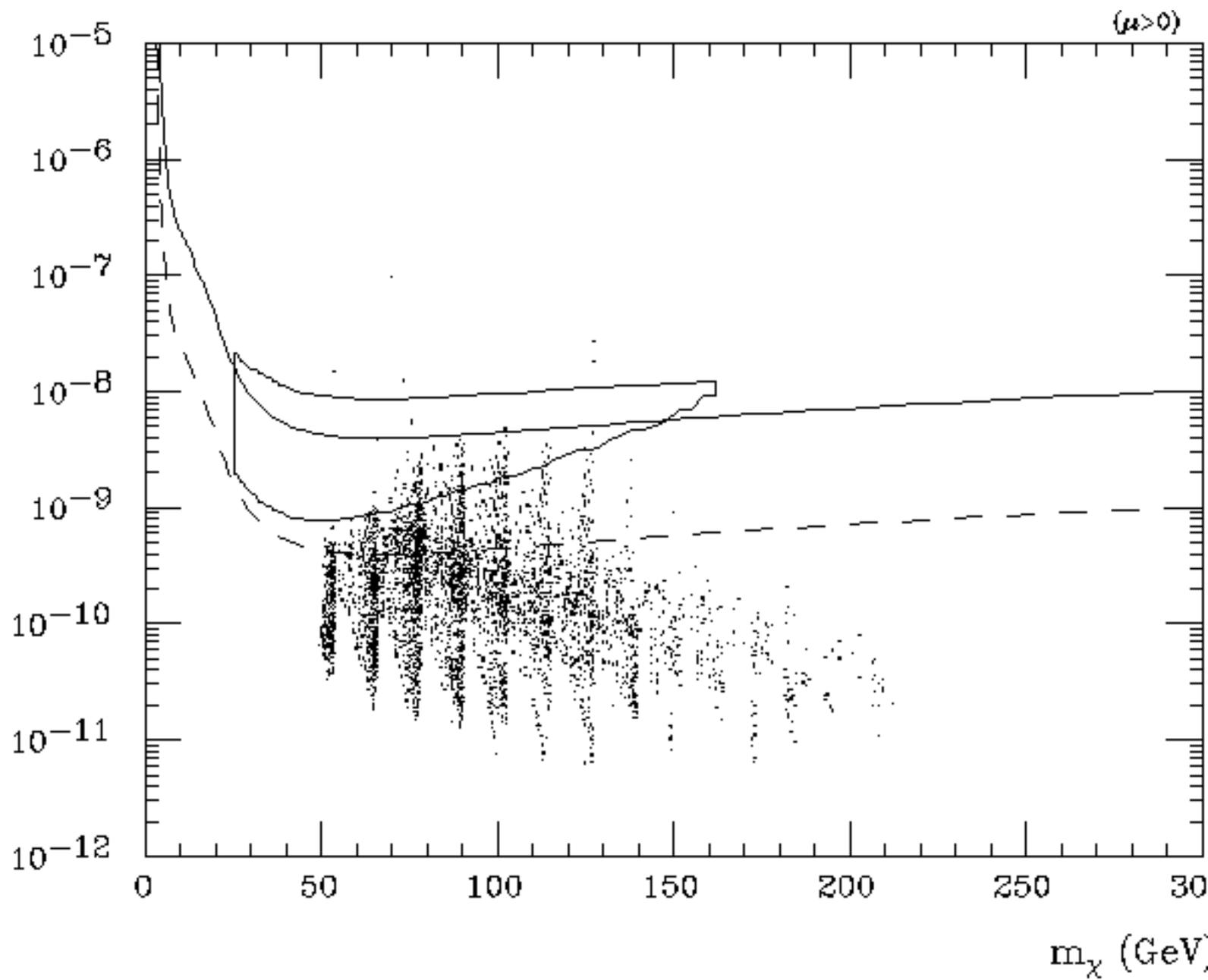
{\bf Fig. 15} -  Scatter plot of the values of 
$\xi \sigma_{\rm scalar}^{\rm nucleon}$ versus the neutralino mass for 
the configurations of set $M$ (a) and of set
$N$ (b). The open curve denotes the (90 \% C.L.) upper bound obtained
from experimental data of Ref. \cite{few}. The region delimited by a closed 
contour is the
one singled out by the experiment of Ref. \cite{dama} as possibly indicative of
an annual modulation effect. The total local dark matter density is normalized
here to the value $\rho_l = 0.4$ GeV cm$^{-3}$. 
The dashed line shows
the discovery potential in case of an improvement by a factor of 10 in current
sensitivities for experiments of direct search for particle dark matter. 

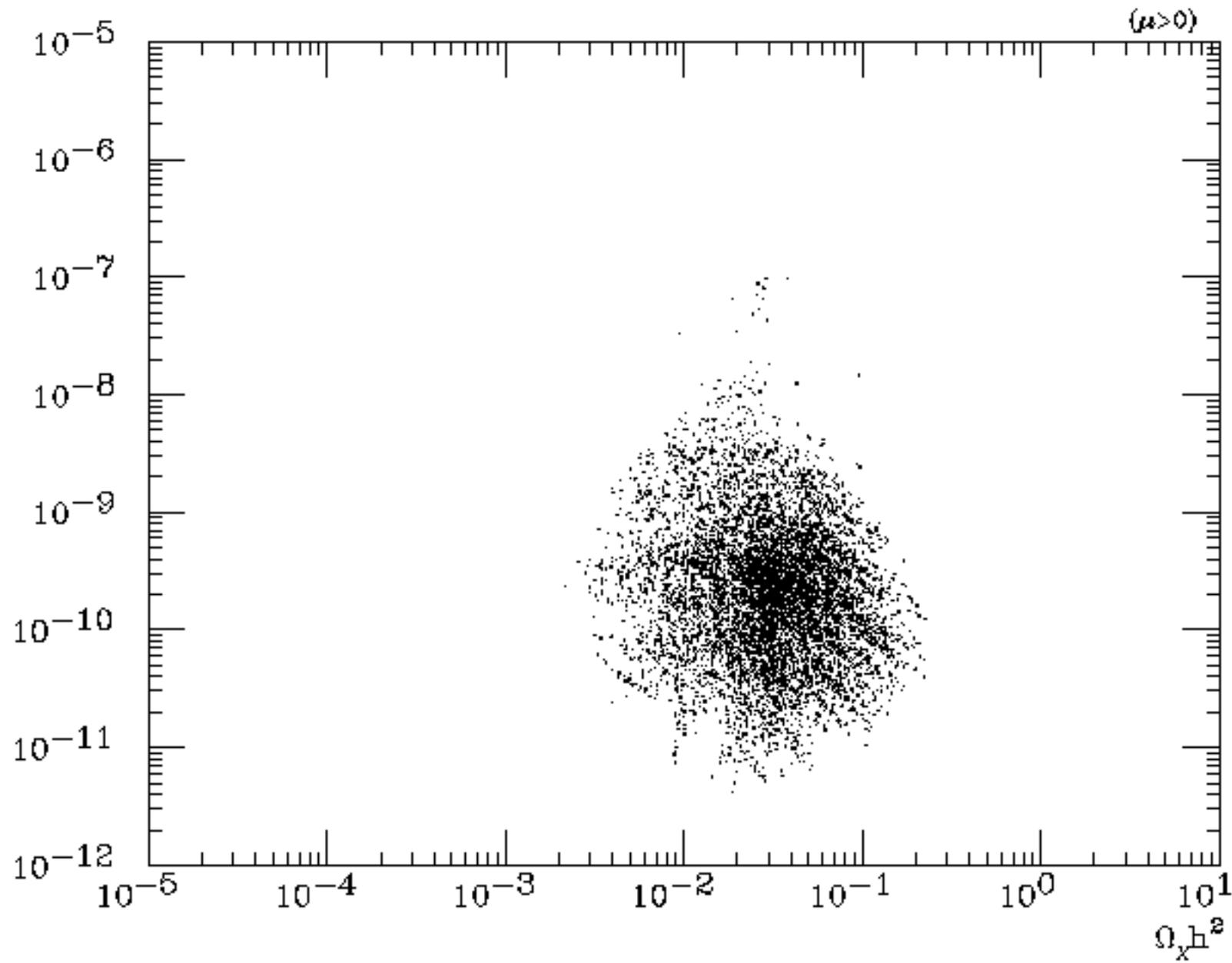
{\bf Fig. 16} - Correlation between $\xi \sigma_{\rm scalar}^{\rm nucleon}$
and the neutralino relic density $\Omega_{\chi} h^2$ 
for configurations of set $M$. 

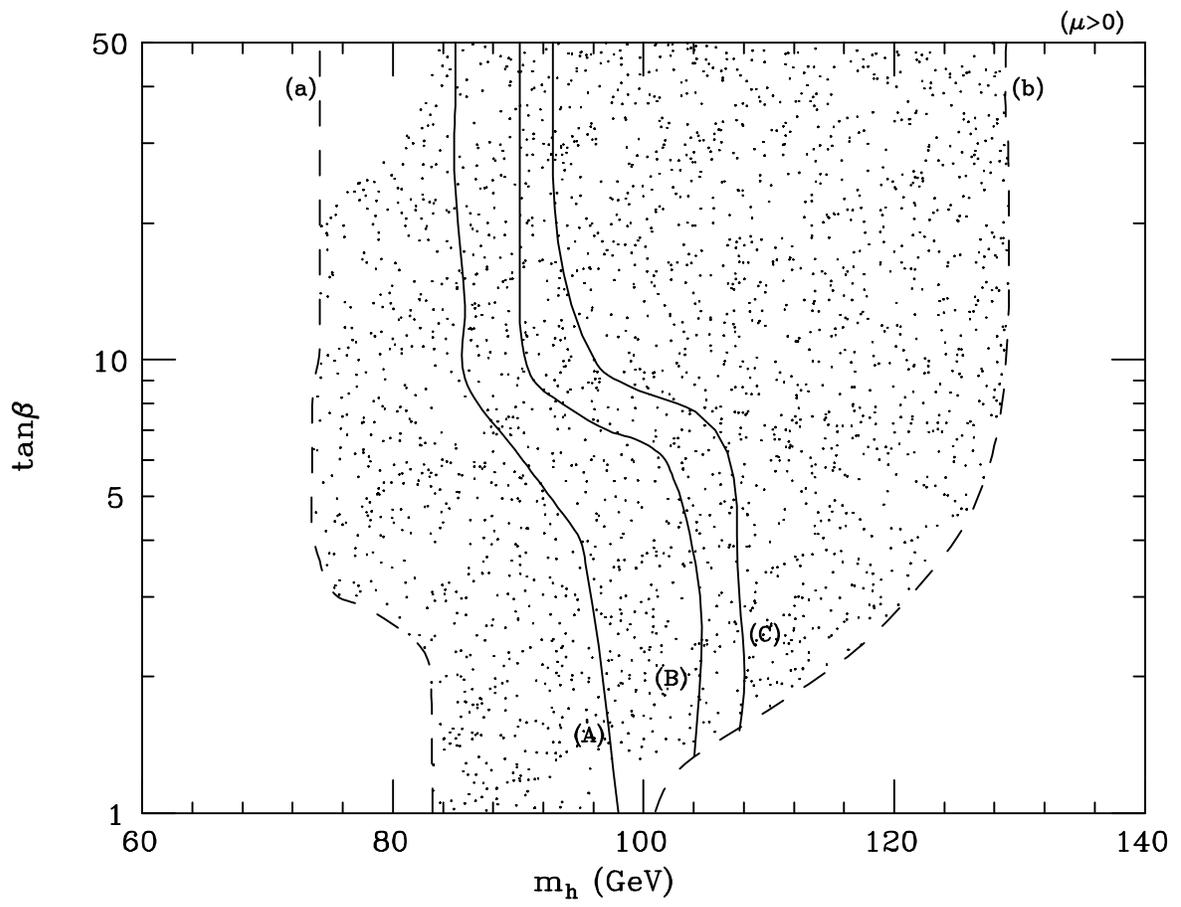
{\bf Fig. 17} - Scatter plot for configurations of set $M$ in the 
$m_h$--$\tan \beta$ plane. 
The region on the left of the dashed line denoted by (a) is excluded by current
LEP experimental data \cite{lep182}, the one on the 
right of the dashed line (b) is theoretically disallowed.
The other lines display the LEP reach at luminosity $L = 200$ pb$^{-1}$ and
various energies \cite{alt}: (A) discovery potential at ${\sqrt s} = 192$ GeV;
(B) discovery potential at ${\sqrt s} = 200$ GeV; 
(C) exclusion at ${\sqrt s} = 200$ GeV.

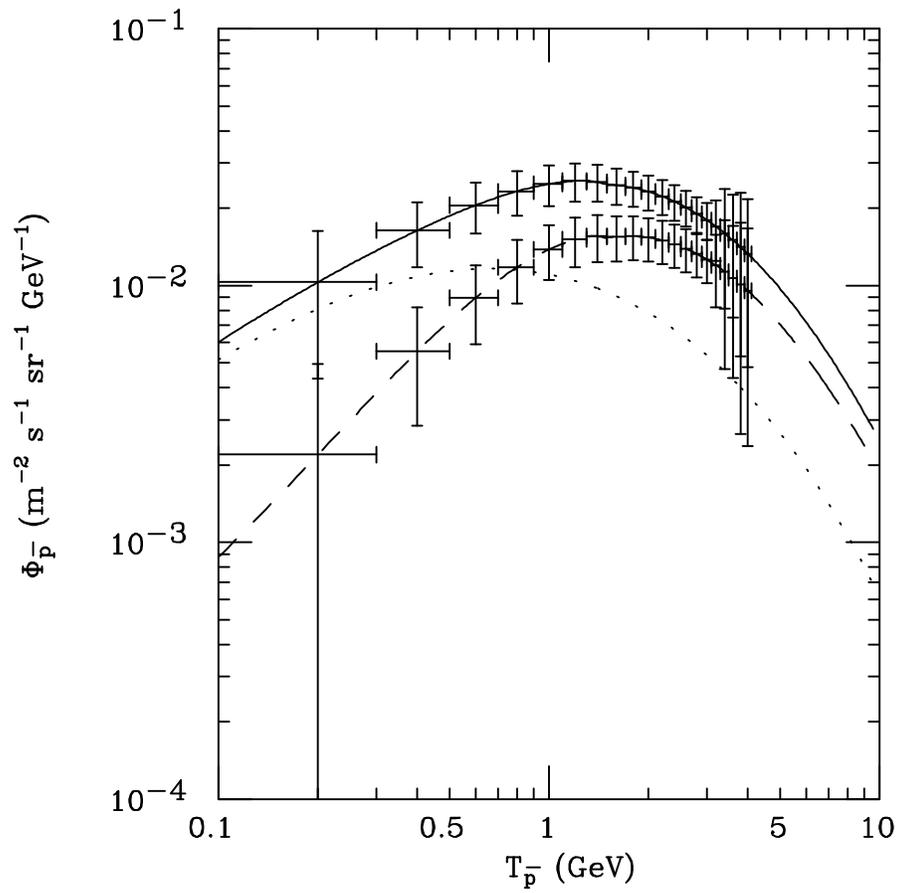
{\bf  Fig. 18} - Expected distribution of measurements with the AMS 
Shuttle flight according to
two different hypothesis: a) dominance of the secondary contribution
(lower sequence  of crosses), b) significant contribution due a
neutralino--induced signal (upper sequence of crosses). 
The dashed line denotes the secondary flux, the dotted one 
denotes the primary flux due to neutralino annihilation in the halo for a
neutralino configuration with the representative values: 
$m_{\chi} = 62$ GeV, $P = 0.98$ and $\Omega_{\chi} h^2 = 0.11$.
Solid line denotes the calculated total flux.

\vfill

\begin{thebibliography}{99}
\bibitem{note1} A possible interpretation of BESS95 data in terms of
neutralino  annihilation is also suggested in Ref. \cite{bess95}. However the
theoretical models used in that analysis (taken from Ref.\cite{mitsui})
differ from ours both for primary and
secondary fluxes. In particular, the supersymmetric model employed there 
is a minimal N =1 supergravity model, at variance with the one used in the 
present paper (MSSM at the electroweak scale). Furthermore, a large enhancement
effect in the primary flux (induced by clumpiness in the halo) is introduced
in that analysis.
\bibitem{note} These data are consistent with the results of two other recent
experiments \cite{mitchell,barbiellini}.
\bibitem{chi2} Since the BESS95 data have asymmetric error bars, we define a
generalized $\chi^2$ as:
\begin{displaymath}
\chi^2 = \sum_i \left\{ \frac{(\Phi^{\rm th}(T_i)-\Phi^{\rm exp}_i)^2}
{(\sigma_i^+)^2}\,
\theta(\Phi^{\rm th}(T_i)-\Phi^{\rm exp}_i)+ 
\frac{(\Phi^{\rm th}(T_i)-\Phi^{\rm exp}_i)^2}{(\sigma_i^-)^2}\,
\theta(\Phi^{\rm exp}_i-\Phi^{\rm th}(T_i)) \right\}, \nonumber
\end{displaymath}
where $\sigma_i^+$ ($\sigma_i^-$) is the positive (negative)
uncertainty in the i--th energy bin and $\theta$ is the Heaviside step
function. 
In the case of symmetric uncertainties 
($\sigma_i^+ = \sigma_i^- \equiv \sigma_i$) 
this formula gives the standard definition of $\chi^2$.
\end{thebibliography}
\end{document}